%% file: main.tex
\documentclass[journal]{vgtc}

\input{config/config}

\onlineid{1528}

\vgtccategory{ISMAR 2026 Papers}

\title{\sysname{}: Mixed Reality Search-and-Rescue\\ Platform for Cognitive Attack Evaluation}

\author{%
  \authororcid{Mingyu Zhu}{0009-0005-2672-7819},
  \authororcid{Jiangong Chen}{0000-0002-1890-5110},
  \authororcid{Yu Liu}{0000-0003-4379-128X},
  \authororcid{Hillmer Chona}{0009-0005-6437-5842},
  \authororcid{Zifan Zhou}{0009-0006-7067-8353},
  \authororcid{Jianan Jiang}{0000-0003-3029-8146},\\
  Jonathan Dodge,
  \authororcid{Feng Qian}{0000-0001-8509-2650},
  \authororcid{Mahdi Imani}{0000-0001-9570-9909},
  \authororcid{Tian Lan}{0000-0003-3010-8090},
  \authororcid{Gang Tan}{0000-0001-6109-6091},
  and \authororcid{Bin Li}{0000-0001-6002-677X}
}

\authorfooter{
  \item
  Mingyu Zhu, Jiangong Chen, Hillmer Chona, Zifan Zhou, Jianan Jiang,
  Jonathan Dodge, Gang Tan, and Bin Li are with The Pennsylvania State University.
  E-mail: \{mintrrey, jiangong, hac5324, zqz5454, jnjiang, jxd6067, gtan, binli\}@psu.edu.

  \item
  Yu Liu and Feng Qian are with the University of Southern California.
  E-mail: \{yliu2170, fengqian\}@usc.edu.

  \item
  Mahdi Imani is with Northeastern University.
  E-mail: m.imani@northeastern.edu.

  \item
  Tian Lan is with The George Washington University.
  E-mail: tlan@email.gwu.edu.
}

\teaser{
  \centering
  \includegraphics[width=\linewidth]{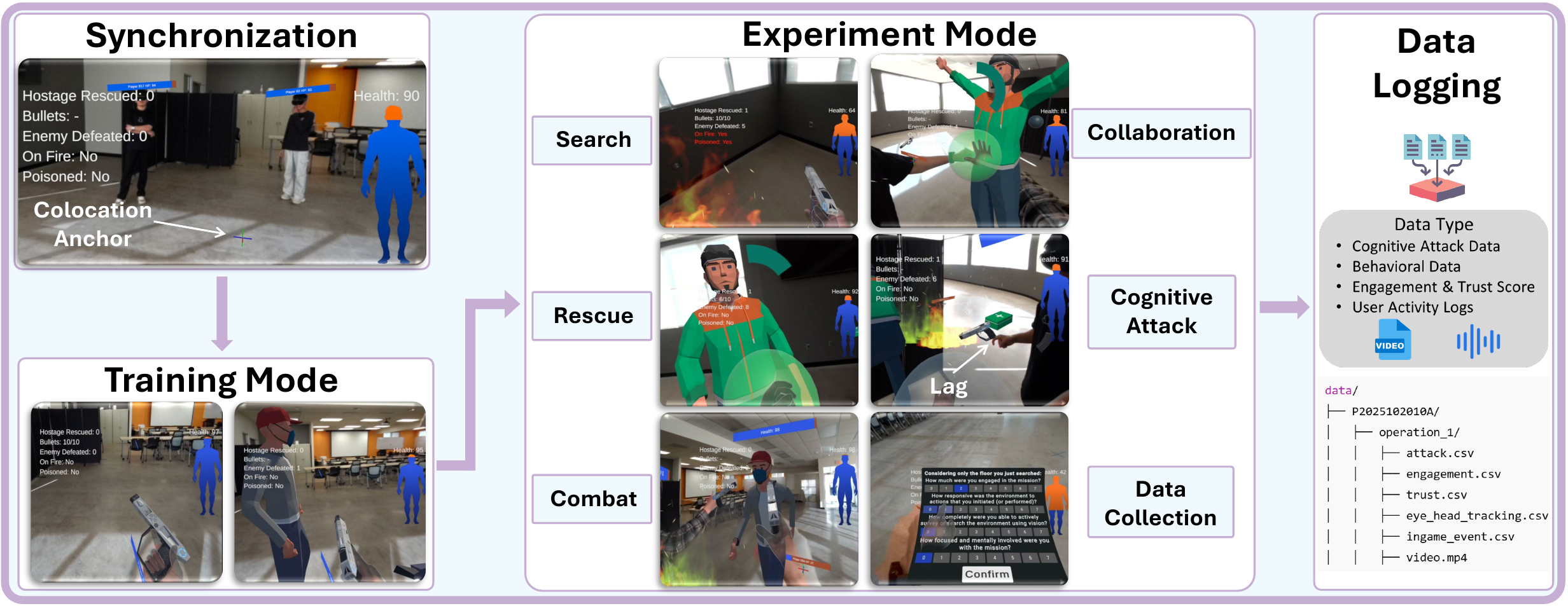}
  \caption{Overview of the mixed reality search-and-rescue testbed. The platform begins with synchronization using a colocation system, followed by a training phase, and then transitions into experimental tasks including search, rescue, and combat. Collaborative interactions, cognitive attacks, and data collection are seamlessly integrated into the testbed, supporting synchronized multi-user scenarios, structured training, and systematic evaluation under cognitive attacks.}
  \label{fig:teaser}
}

\input{section/00-abstract}

\begin{document}

\firstsection{Introduction}

\maketitle

\input{section/01-intro}
\input{section/02-related}
\input{section/03-attack}
\input{section/04-system}
\input{section/05-core_feature}
\input{section/06-multiplayer}
\input{section/07-data}

\input{section/08-user_study}
\input{section/09-discussion}
\input{section/10-conclusion}

\acknowledgments{
This work has been supported by DARPA grant HR0011-2420366.}

\bibliographystyle{config/abbrv-doi}
\bibliography{reference}

\clearpage
\twocolumn[
    \begin{center}
        {\Large\bfseries Supplementary Material}
    \end{center}
    \vspace{1em}
]

\appendix
\crefalias{section}{appendix}

\input{SupplementaryMaterial}

\end{document}

%% file: config/config.tex
\graphicspath{{figures/}{pictures/}{images/}{./}} 

\usepackage{times}                     

\usepackage{tabu}                      
\usepackage{booktabs}                  
\usepackage{lipsum}                    
\usepackage{mwe}                       

\usepackage{mathptmx}
\usepackage[caption=false]{subfig}

\usepackage{tabularx}
\usepackage{array} 
\newcolumntype{L}{>{\raggedright\arraybackslash}X}

\usepackage{enumitem}

\newcommand{\sysname}[1]{{CAESAR}}

%% file: section/00-abstract.tex
\abstract{
Mixed Reality (MR) enables immersive training by tightly coupling virtual content with the physical environment and supporting embodied interaction.
In safety-critical collaborative tasks such as search-and-rescue, however, users rely heavily on virtual elements for perception, coordination, and decision-making, making them vulnerable to cognitive attacks that subtly distort visual feedback, virtual content, or system synchronization.
While prior MR security research has examined individual attack mechanisms, evaluations are typically conducted in isolated or single-user settings, and there is a lack of standardized collaborative testbeds for systematic adversarial experimentation.
In this paper, we develop \sysname{}, an end-to-end multi-user MR framework for systematic evaluation of cognitive attacks in collaborative search-and-rescue training.
\sysname{} features a modular architecture that integrates four classes of cognitive attacks within a synchronized collaborative scenario and incorporates a multimodal data collection pipeline that captures behavioral, performance, and subjective measures under adversarial conditions.
We conduct a controlled user study with an in-depth analysis of the latency-based attack, demonstrating measurable effects on task performance and user behavior while characterizing users' subjective experience under adversarial conditions.
}

\keywords{Mixed reality, cognitive attack, search and rescue, evaluation testbed}

%% file: section/01-intro.tex
\label{sec:introduction}

Mixed Reality (MR) enables immersive training experiences that tightly couple virtual content with the physical environment, allowing users to engage in embodied interactions with both digital and real-world elements~\cite{ersozlu2021mixed, frank2017mixed, pellas2020systematic, duong2024mrclassroom} while gaining hands-on experience that is critical for effective skill acquisition~\cite{azimi_can_2018, dong2023collaborative, gonzalez2017immersive}.
Unlike Virtual Reality (VR), which fully replaces the physical world with synthetic simulations~\cite{lorenz2025hips, zhao2020learning}, MR maintains physical grounding while augmenting perception and interaction, enabling training that more closely reflects real-world operations.
These capabilities are particularly valuable in safety-critical domains, where real-world training may be dangerous or infeasible, such as firefighter training~\cite{dong2023collaborative, safranoglou_augmented_2024}, emergency medical procedures~\cite{azimi_can_2018}, and collaborative search-and-rescue operations~\cite{li2022theoryOfMindSAR, Army2020, willet2023teamCogLoad}.

In collaborative search-and-rescue MR training, virtual elements directly shape users’ attention, perception, and coordinated action across shared tasks.
Because decision-making depends on accurate spatial alignment, synchronized state representation, and interpretation of virtual content, even subtle perturbations can influence task execution and collective situational awareness.
This tight coupling between perception, interaction, and system infrastructure introduces a vulnerability: perturbations to visual feedback, virtual content, user-state information, or networked synchronization can distort cognition without immediately alerting users.
Despite this risk, existing search-and-rescue MR systems primarily aim to enhance operational effectiveness, realism, and situational awareness. For example, prior work has explored agent-based MR simulation for combat search-and-rescue command and control~\cite{ji2019agent}, collaborative perception systems~\cite{wilchek2025ajna}, and Augmented Reality (AR)-based urban search-and-rescue support integrating AI and localization~\cite{wilchek_khait_2025}. However, these systems are not designed to support controlled injection of adversarial perturbations or systematic evaluation of user responses under degraded or manipulated conditions.

Prior MR security research~\cite{cheng2023perceptual, sajid2025just, xiu2025detecting} has examined individual attack mechanisms, but typically in isolated or single-user contexts.
As a result, there remains a lack of standardized collaborative, end-to-end MR testbeds that enable systematic evaluation of cognitive attacks within realistic multi-user training scenarios.
In particular, few platforms capture synchronized behavioral, decision-making, and cognitive data under adversarial conditions.
Such data are essential for understanding user behavior, performance, and cognitive responses and for advancing cognitive security research~\cite{oppelt2022adabase}.

To address this gap, we develop CAESAR (\underline{C}ognitive \underline{A}ttack \underline{E}valuation in \underline{S}earch-\underline{A}nd-\underline{R}escue), an end-to-end collaborative MR experimental framework and testbed for systematic evaluation of cognitive attacks in multi-user search-and-rescue scenarios.
\sysname{} integrates four classes of cognitive attacks, including MR display perturbation, MR content manipulation, MR user status breach, and network and compute perturbation, within a synchronized end-to-end scenario.
Integrated with a multimodal data collection pipeline that records attack events, behavioral signals, task performance, and subjective measures, the testbed enables comprehensive analysis of user behavior and collaborative dynamics under adversarial conditions.

The \sysname{} testbed models a realistic multi-user search-and-rescue mission in which participants locate and rescue hostages within a hazardous and adversarial MR environment.
The scenario incorporates a unified health system that introduces time pressure and risk-reward tradeoffs, dynamic environmental hazards that degrade perception and safety, and AI-driven adversaries that impose continuous external threats.
Hostage rescue requires sustained interaction and coordination, coupling combat and rescue objectives through interdependent task mechanics.
The platform maintains fully synchronized multi-user collaboration to preserve shared situational awareness and coordinated action.
To scale mission complexity within constrained physical spaces, \sysname{} employs a virtual elevator mechanism that structures the experience into multi-stage missions across distinct virtual environments.

These design elements establish \sysname{} as an end-to-end collaborative MR testbed for systematic evaluation of cognitive attacks in search-and-rescue training. Our contributions are as follows:
\begin{itemize}[nosep]
    \item We develop and open-source\footnote{\url{https://github.com/SNeC-Lab-PSU/CAESAR}}, to our knowledge, the first end-to-end multi-user MR testbed designed for systematic evaluation of cognitive attacks in collaborative search-and-rescue scenarios.
    \item We implement four classes of cognitive attacks within the testbed and conduct an in-depth user study that uses latency-based display perturbation to validate the framework, demonstrating its impact on task performance, user behavior, and subjective experience.
    \item We provide an integrated experimental framework that includes configurable gameplay modes, controlled attack injection, and a real-time multimodal data collection pipeline, enabling systematic user studies and generating rich datasets for post-hoc analysis, and supporting future collaborative MR cognitive attack research across diverse application scenarios.
\end{itemize}

%% file: section/02-related.tex
\section{Related Work}\label{sec:related-work}

\subsection{Security Attacks and Recognizability in MR}
\label{sec:related_MRsecurity}

The integration of real-world environments with virtual content in MR systems introduces novel security risks and attack surfaces.
Attackers can manipulate visual information in MR to present misleading content, potentially causing users to make unsafe or erroneous decisions.
Virtual objects may also occlude real-world elements or distort spatial relationships, further increasing the risk of user error.
Recent work has begun to formalize the broader landscape of MR cognitive and deception attacks. For example, Teymourian et al.~\cite{teymourian2025sok} proposed a comprehensive ontology of MR deception attacks spanning security, cognition, and information-theoretic perspectives.
Alongside this emerging understanding, prior research has investigated different types of attacks in MR and AR systems, with a focus on their recognizability by users.
Cheng et al.~\cite{cheng2023perceptual} conducted an early empirical study of perceptual manipulation attacks in MR, showing that users often struggle to attribute anomalous perceptions to malicious interference.
Subsequent work by Cheng et al.~\cite{cheng2024ui} demonstrated that embedding users within AR interfaces undermines traditional UI security assumptions, making attacks harder to detect and reason about.
Other studies~\cite{xiu2025detecting, xiu2024lobstar, xiu2025viddar} examined visual information attacks such as misleading overlays, occlusion of critical elements, and misaligned virtual content.
Sajid et al.~\cite{sajid2025just} explored interaction attacks in MR and found that manipulations of spatial cues are particularly difficult for users to recognize.

While these studies provide valuable insights into individual attack mechanisms, they are typically evaluated in isolation and task-specific or single-user settings.
As a result, they offer limited insight into how MR security threats manifest in realistic, collaborative environments.
As multi-user MR applications become increasingly common, how attacks affect shared spatial understanding, coordination, and collective decision-making remains underexplored.

\subsection{Co-located Multi-user Collaboration in MR}
\label{sec:related_multiuser}

Many MR applications, especially those for training, planning, and decision support, are inherently collaborative, which fundamentally alters how security threats manifest and propagate across users.
MR systems support both co-located~\cite{chen2026prism, hu2023mofa, jansen2020share} and remote~\cite{gronbaek2024blended, sidenmark2024desk2desk, yao2022scalable} collaboration, enabling users to share virtual content while maintaining awareness of the physical environment and others' actions.
For co-located users, MR facilitates situated and embodied collaboration by anchoring virtual objects to a shared physical space.
Prior work has shown that effective multi-user MR collaboration depends on maintaining a common spatial reference frame and supporting awareness of collaborators’ activities~\cite{augello2023design, ens2019revisiting}.
Several systems have been proposed to enhance shared understanding and interaction consistency.
For example, Gottsacker et al.~\cite{gottsacker2025decoupled} presented a decoupled rendering approach that improves communication by visualizing collaborators’ actions from their respective viewpoints, while Miria~\cite{buschel2021miria} provides a co-located MR data visualization toolkit based on manual coordinate synchronization.

Although these systems improve usability and scalability, they are primarily designed to enhance interaction quality rather than assess system robustness or security.
From a security perspective, inconsistencies in synchronization, misaligned spatial representations, or asymmetric visual states across users can enable attacks that are difficult to detect individually.
However, few existing collaborative MR frameworks support systematic evaluation of such vulnerabilities under adversarial or degraded conditions.

\subsection{Simulation and Training in MR}
\label{sec:related_MRTraining}
MR and VR technologies have been widely adopted for simulation and training because they provide immersive, interactive experiences while reducing the cost and risk of real-world training.
They have been applied across domains such as education, medical training, firefighting, and search and rescue.
For example, Zhao et al.~\cite{zhao2020learning} demonstrated that VR video and pre-captured scenes can substitute for physical field trips in geometry education, while Lorenz et al.~\cite{lorenz2025hips} incorporated haptic feedback into VR-based hip surgery training to improve procedural realism.
Compared with VR, MR enables training directly within real-world environments while augmenting perception and interaction, making it particularly suitable for safety-critical, team-based tasks.
For example, Dong et al.~\cite{dong2023collaborative} introduced a collaborative MR system for firefighter training that emphasizes synchronized multi-user experiences.
Search and rescue operations are similarly safety-critical and inherently collaborative, involving complex social and cognitive dynamics~\cite{li2022theoryOfMindSAR, Army2020, willet2023teamCogLoad}. Ji et al.~\cite{ji2019agent} presented an early agent-based MR simulation framework for combat search and rescue command and control, while Ajna~\cite{wilchek2025ajna} improves situational awareness in search and rescue through shared perception, focusing on operational effectiveness rather than security or adversarial robustness.
Related work has also explored AR-based approaches for urban search and rescue, such as integrating AI and localization to support human-canine teams~\cite{wilchek_khait_2025}.
Beyond application-specific training systems, research platforms such as ILLIXR~\cite{huzaifa2021illixr} provide modular XR infrastructure for end-to-end systems experimentation. Rather than targeting a specific application domain, these platforms support XR systems research and development.

Despite these advances, prior MR and VR training systems primarily aim to improve realism, usability, and task performance.
Few are designed as experimental testbeds for evaluating security threats, and they do not support controlled injection of cognitive attacks, systematic variation of attack parameters, or unified measurement of system- and user-level responses in multi-user settings. As a result, how collaborative MR training systems and their users behave under security threats remains poorly understood.
In contrast, we present an end-to-end collaborative MR experimental framework designed as a testbed for systematic evaluation of cognitive attacks, integrating configurable attack injection, synchronized multi-user experimentation, and multimodal behavioral measurement within collaborative search-and-rescue scenarios.
By doing so, we bridge the gap between MR security research and MR-based training systems, contributing toward safer and more resilient multi-user MR platforms.

%% file: section/03-attack.tex
\section{Cognitive Attack} \label{sec:cog_attack}

Modern MR systems, while enabling immersive collaboration, also introduce opportunities for adversarial manipulation of user cognition.
We define cognitive attacks as intentional disruptions that exploit MR’s access to sensory, perceptual, and behavioral modalities to overload, mislead, or destabilize users during task execution. In this section, we present the cognitive attack classes implemented within our framework for systematic evaluation in collaborative MR environments. These attacks leverage critical MR system dependencies, such as visual feedback, sensor data, or network and compute resources, to expose potential vulnerabilities and their impact on user cognition. \Cref{fig:cog-attack} shows an overview of the attack classes considered in this work.
\input{figure/02-attacks}

\subsection{Threat Model and Design Rationale}
\textbf{Threat Model.}
Our objective is to study the effects of cognitive attacks on user behavior and task performance, rather than the feasibility of compromising MR platforms or bypassing platform security mechanisms. Accordingly, we consider an adversary whose goal is to manipulate user cognition in MR environments by degrading perception, misleading interpretation, or increasing cognitive load during task execution, rather than causing system failure.
The attacker does not require physical access to the victim but can influence the MR system through software, networking, or content-level channels.
Such an attacker may take the form of a compromised MR application or plugin~\cite{cheng2024ui}, a malicious or misconfigured server in a client--server architecture~\cite{sajid2025just}, or a third-party service with access to MR system APIs~\cite{cheng2023perceptual}. We assume the attacker cannot bypass hardware-level isolation or the underlying operating system but operates at the application or middleware level by exploiting exposed interfaces and trusted execution paths commonly used in collaborative MR systems.
Under this model, the attacker can modify or delay application-level events, alter virtual content or feedback logic, access user sensing streams that are legitimately exposed through platform APIs or granted permissions, and interfere with network synchronization or compute resource allocation.

\textbf{Design Rationale.}
Modern MR experiences rely on a small set of critical system dependencies that directly shape user cognition, including (1) continuous visual feedback and action--feedback coupling, (2) semantically meaningful virtual content and interaction logic, (3) rich user sensing pipelines that capture behavioral signals, and (4) networked and compute-intensive execution for multi-user synchronization and real-time processing.
Attacks that target any of these dependencies can systematically disrupt perception, attention, trust, or decision-making without immediately alerting the user.

Based on this observation, we organize the cognitive attacks into four categories corresponding to these core MR system dependencies: MR display perturbation, MR content manipulation, MR user status breach, and network and compute perturbation.
Unlike ontology-oriented frameworks that seek to comprehensively characterize MR deception attacks~\cite{teymourian2025sok}, our implementation-oriented categorization is designed to support systematic deployment and evaluation within collaborative MR systems instead of proposing a comprehensive attack taxonomy.
Rather than enumerating all exploits, these categories reflect distinct security-relevant MR system layers identified in prior work~\cite{el2024cybersecurity,garrido2023sok, happa2019cyber}, capturing common and high-impact cognitive attack surfaces.
They provide structured coverage of adversarial manipulation across perception, interaction logic, sensing pipelines, and infrastructure, motivating the implemented attack classes below.

\subsection{Implemented Cognitive Attack Classes}

\textbf{MR Display Perturbation.}
MR display perturbation represents a class of attacks that target continuous visual feedback and action--feedback coupling in MR environments.
An attacker operating at the application or rendering layer can intentionally degrade visual fidelity, responsiveness, or tracking stability without causing overt system failure.
Such manipulations disrupt the tight perception--action loop, forcing users to adapt under degraded conditions and increasing cognitive workload.
The following attack instances demonstrate practical realizations of display perturbation in MR systems and are implemented in our testbed.

\emph{Frame Rate Reduction.}
Artificial delays can be introduced into the update or rendering loop, reducing the effective frame rate perceived by the user.
Such degradation can be achieved through inserted wait states, blocking operations, or compute-intensive callbacks.
In our testbed, we implement this attack by injecting controlled delays into the main update loop.
Reduced frame rates disrupt visual smoothness and weaken action--feedback coupling, leading to delayed interaction.

\emph{Action-to-Display Latency.} 
Delays can be introduced between user input events and corresponding visual updates, disrupting the temporal consistency of the action--feedback loop.
This can be realized by intercepting or delaying event handling callbacks at the application layer.
In our testbed, we implement this attack by inserting controlled delays and random jitter into input processing before corresponding rendering updates are triggered.
The resulting action--response mismatch degrades responsiveness and destabilizes perceived control.

\emph{Brightness and Color Alteration.}
Display-level parameters exposed through platform APIs can be manipulated to alter the visual appearance of the physical environment in MR.
Using platform-provided passthrough control interfaces (e.g., Meta's Passthrough API~\cite{metapassthrough}), brightness and edge color rendering can be adjusted without privileged system access.
In our testbed, we implement this attack by dynamically adjusting passthrough brightness and edge color values during task execution.
Such alterations distort scene visibility and saliency, potentially misleading perception and increasing cognitive workload.

\emph{Camera Tracking Degradation.}
Tracking stability can be degraded by introducing controlled noise or offsets into pose data consumed at the application layer.
Such perturbations can be applied to camera position and orientation values obtained through platform APIs, without modifying hardware or underlying system components.
In our testbed, we implement this attack by injecting stochastic perturbations into the tracked pose data before scene rendering.
These disturbances cause scene jitter or drift, weakening spatial alignment and destabilizing perception.

\textbf{MR Content Manipulation.}
MR content manipulation represents a class of attacks that target the semantic and interaction logic of virtual objects within the MR environment.
An attacker operating at the application layer can modify object properties, interaction handlers, or feedback messages, thereby altering how virtual content is presented and interpreted without causing system failure.
Inconsistencies between appearance, behavior, and task semantics can mislead users and degrade decision-making.
We implement two content-level attack instances in our testbed, as described below.

\emph{False Information Injection.}
Interaction feedback logic can be altered to provide incorrect or misleading semantic information by modifying system messages or status indicators, without changing the underlying interaction state.
In our testbed, we modify the feedback handler associated with a hazardous virtual object to display incorrect information despite its actual behavior.
This mismatch between system feedback and environmental state can mislead users and undermine situational awareness.

\emph{Visual Object Manipulation.}
The spatial alignment between a virtual object’s visual representation and its interaction boundary can be deliberately altered.
By modifying object transforms or interaction colliders, the rendered object can be displaced from its true interaction location.
In our testbed, we shift the visual representation of a hazardous object while keeping its original interaction boundary unchanged.
This visual--interaction mismatch disrupts perception--action mapping and may confuse users.

\textbf{MR User Status Breach.}
MR user status breach represents a class of attacks that exploit sensor data streams exposed to applications in MR systems.
An attacker operating at the application layer can access and log behavioral signals such as head motion and eye gaze without modifying hardware or system components.
Such access enables the extraction of sensitive cognitive and behavioral information.
Moreover, eye gaze and head motion patterns may allow user identification based on behavioral biometrics~\cite{10464194, 9756745}.
In our testbed, we capture real-time gaze and head tracking data, demonstrating that these sensor streams are directly accessible and recordable at the application layer.

\emph{Head and Eye Tracking Visualization.}
The headset continuously exposes head pose and eye tracking data to applications through platform-provided APIs and granted permissions, including 3D position, orientation, and gaze vectors.
In our testbed, we access these streams in real time, compute the gaze focal point, and transmit the collected data to a server for subsequent visualization and analysis.
This demonstrates that detailed behavioral signals can be extracted and externally processed, potentially enabling the types of inference and identification risks.

\textbf{Network and Compute Perturbation.}
Network and compute perturbation represents a class of attacks that target MR systems’ dependence on communication latency and computational resources.
An attacker operating at the application or server layer can intentionally introduce delays, inject artificial workload, or manipulate synchronization logic.
Such perturbations degrade responsiveness, disrupt multi-user consistency, and weaken temporal coordination in collaborative MR environments.
We implement two network- and compute-level attack instances in our testbed.

\emph{Network Latency Injection.}
Collaborative MR systems rely on timely synchronization of shared virtual state across devices.
Synchronization messages can be intentionally delayed to introduce communication latency.
In our testbed, we modify the server-side synchronization process of shared objects (e.g., via Meta Shared Spatial Anchors API~\cite{metassa}) to introduce controlled delays between object creation and remote rendering.
These induced delays disrupt temporal consistency and degrade the responsiveness and coordination of multi-user MR interactions.

\emph{Compute Load Perturbation.}
Compute load perturbation exploits the limited processing resources of MR applications.
Adversarial workload can be introduced at the application or middleware layer by triggering computation-intensive routines in response to specific events.
In our testbed, we inject workload into the execution pipeline to increase CPU usage and reduce rendering responsiveness.
The resulting resource contention lowers frame stability and degrades interaction smoothness.

%% file: figure/02-attacks.tex
\begin{figure}[tb]
    \centering
    \includegraphics[width=0.47\textwidth]{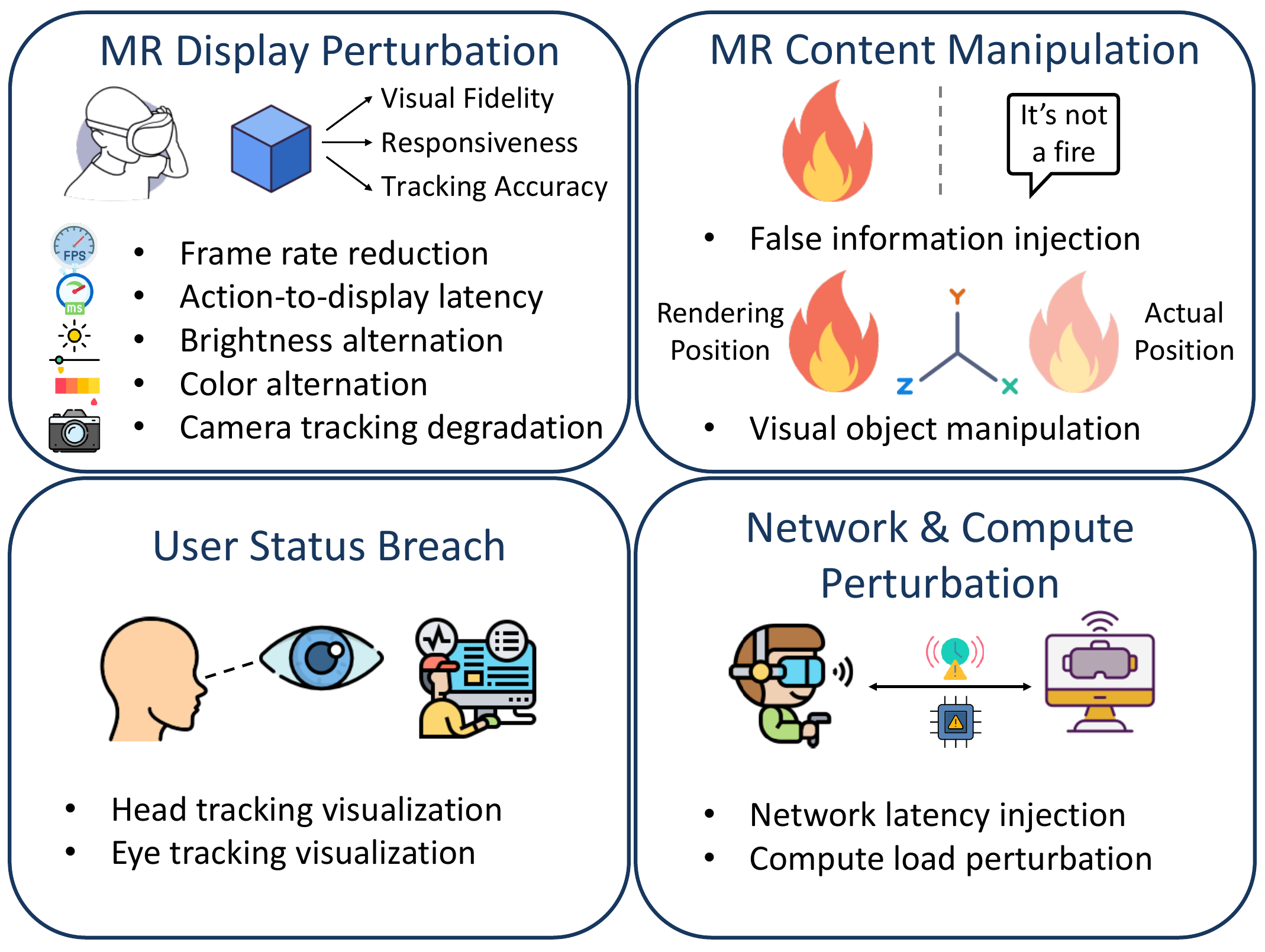}
     \caption{Four classes of cognitive attacks in MR environments.}
    \label{fig:cog-attack}
    \vspace{-0.2in}
\end{figure}

%% file: section/04-system.tex
\section{System Overview}\label{sec:system}

\input{figure/03-architecture}

We designed the \sysname{} framework as a multi-user MR testbed for controlled tactical search-and-rescue simulations with integrated support for cognitive attack deployment.
As shown in~\Cref{fig:sys-arch}, it combines spatially mapped physical environments with persistent virtual objects, synchronized interactions and animations, real-time player-state tracking, and comprehensive data logging.

To bridge physical space and the virtual environment, \sysname{} aligns virtual content to the mapped real world using a standard MR spatial mapping pipeline~\cite{metamruk}.
This alignment enables users to interact naturally with both virtual and physical elements, providing a coherent foundation for immersive simulation.
The testbed instantiates mission-critical virtual entities and environmental hazards, which are spatially distributed to form structured search-and-rescue tasks. 
These configurations integrate operational objectives with environmental constraints to create a realistic collaborative setting. 
Importantly, the system provides runtime control mechanisms that allow cognitive attacks to be injected during task execution, enabling systematic evaluation of user behavior and performance under adversarial MR conditions.

Within the MR environment, \sysname{} maintains consistent state synchronization across all participants to support seamless multi-user collaboration.
A colocation system based on a shared spatial reference frame~\cite{metassa} ensures geometric alignment of virtual content across devices, while real-time network synchronization~\cite{photonfusion} propagates object states, user interactions, animations, and player status information.
These mechanisms preserve shared situational awareness and ensure that mission-critical events remain consistent across perspectives.

To support controlled experimental studies, the \sysname{} testbed provides a dedicated training mode and configurable experiment modes. 
The training mode introduces core tasks and interaction mechanisms to familiarize participants with mission procedures. 
Experiment modes vary task parameters, object distributions, and attack configurations, and allow dynamic runtime switching and cognitive attack injection.
The platform also integrates a real-time data collection pipeline that records both system-level and user-level metrics. 
Captured data includes cognitive attack events, behavioral signals (e.g., head and eye tracking), user activity logs, and subjective measures such as engagement and trust. 
All data are synchronized and stored on the server for subsequent analysis, enabling fine-grained post-hoc evaluation of user resilience, decision-making, and collaboration under adversarial MR conditions.
These components establish \sysname{} as a flexible and extensible research framework for systematic evaluation of cognitive attacks in collaborative MR search-and-rescue environments.

%% file: figure/03-architecture.tex
\begin{figure}[tb]
    \centering
    \includegraphics[width=\columnwidth]{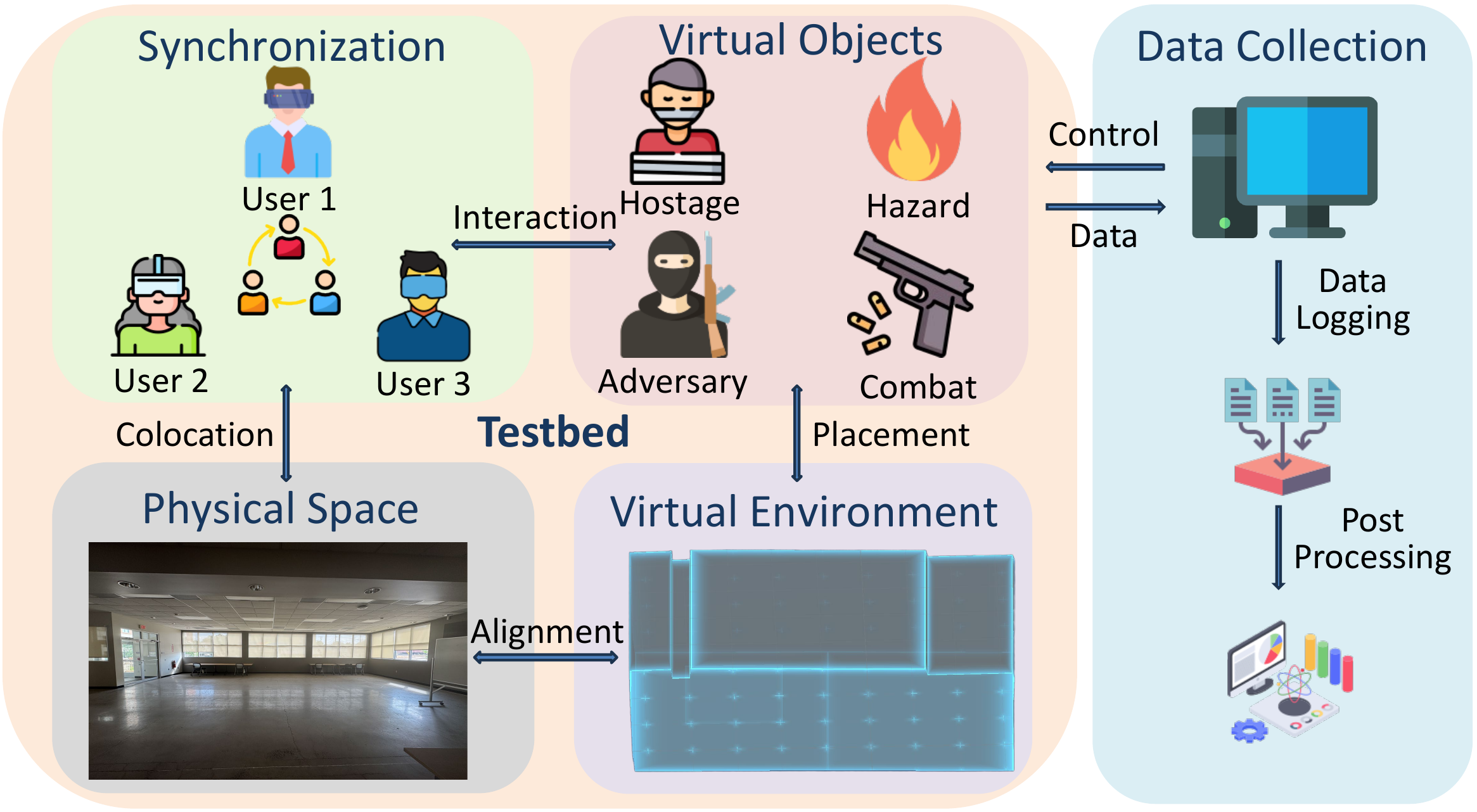}
     \caption{System architecture of the \sysname{} testbed. The framework integrates the physical space with the virtual environment through spatial alignment, and synchronizes multiple users via a colocation system. Within the testbed, users interact with virtual objects such as hostages, adversaries, hazards, and weapons to simulate search-and-rescue tasks. A data collection pipeline captures multimodal data for post-hoc analysis.}
    \label{fig:sys-arch}
    \vspace{-0.2in}
\end{figure}

%% file: section/05-core_feature.tex
\section{Search-and-Rescue Features}

This section describes the core search-and-rescue features supported by the \sysname{} testbed and explains how they shape mission dynamics, user behavior, and experimental control. These components capture key elements of real-world search-and-rescue operations, including time pressure, environmental uncertainty, adversarial interaction, and team coordination, enabling systematic study of collaborative performance under stress.

\input{figure/04-SAR_features}
\subsection{Health System}

We introduce a unified health system to model survival constraints, time pressure, and resource dependency within the search-and-rescue scenario. 
Health serves as a shared quantitative state for players, hostages, and adversaries, linking environmental hazards, combat interactions, and rescue urgency within a single framework (see~\Cref{fig:SAR_features} (Left)).
Players experience gradual health decay over time and additional damage from environmental hazards and adversarial encounters. 
Health can only be restored through limited recovery resources, creating strategic trade-offs between exploration, combat, and self-preservation. 
Hostages also experience health degradation, with varying survival durations that require players to prioritize rescue decisions under uncertainty. 
Adversaries possess health states that decrease through combat, enabling bidirectional risk during engagement.
Importantly, health directly constrains interaction capabilities. 
When a player's health is depleted, their ability to perform mission actions is restricted until recovery is achieved. 
This coupling between physiological state and action capability introduces sustained cognitive and behavioral pressure, encouraging risk assessment, coordination, and adaptive decision-making throughout the mission.

\subsection{Hostile Environment}

We model a hostile environment through four hazard categories: fire, poison gas, explosion, and smoke (see~\Cref{fig:SAR_features} (Left)). These hazards introduce spatial risk, stochastic damage, and perceptual degradation, collectively simulating the uncertainty and dynamics of real-world search-and-rescue operations. Fire and gas impose continuous or residual health loss, while explosions create sudden high-impact events that disrupt navigation and decision-making. Smoke reduces visibility and object detectability, degrading situational awareness. These hazards generate dynamic environmental pressure, forcing players to balance exploration, rescue urgency, and personal safety under uncertain and evolving conditions.

\subsection{Rescue Mechanism}
\label{sec:rescue}

The rescue mechanism implements time-sensitive, collaborative interaction under environmental and adversarial pressure (see~\Cref{fig:SAR_features} (Top Right)). Hostages experience gradual health degradation, creating urgency and requiring timely intervention. Successful rescue requires sustained interaction at close range, exposing players to surrounding risks. To promote teamwork, rescue efficiency increases when multiple players coordinate their interaction, encouraging cooperative behavior rather than isolated action. Rescue attempts are further constrained by mission requirements: players must eliminate a specified number of adversaries before a hostage can be saved. The required number increases progressively across rescues, linking combat performance with rescue progress and introducing increasing coordination demands. These mechanisms transform rescue from a simple interaction into a structured task involving prioritization, synchronization, and risk management under evolving mission constraints.

\subsection{Adversary AI and Combat System}
\label{sec:adversary-combat}

Adversaries introduce adaptive threat pressure through autonomous search, detection, pursuit, and attack behaviors (see~\Cref{fig:SAR_features} (Bottom Right)). In the absence of nearby players, adversaries patrol the environment; once a player is detected, they transition to pursuit and cause damage during close-range engagement. This dynamic response creates continuous situational awareness demands and spatial risk for players. Combat interactions are bidirectional: players can defeat adversaries through ranged or close-range engagement, while adversaries can also inflict damage in return. Ranged combat requires ammunition management and timely reloading, introducing resource constraints that influence engagement decisions. Damage exchange is configurable, allowing controlled adjustment of combat intensity for different experimental conditions.
Adaptive adversary behavior and resource-constrained combat establish sustained threat exposure, requiring players to balance combat decisions, positioning, and coordination under ongoing threat.

\subsection{Elevator System and Visibility Mechanics}
\label{sec:elevator}

To structure mission progression and extend the effective study space, we introduce a virtual elevator system that segments the environment into multiple virtual floors. Each floor contains distinct distributions of hostages, adversaries, and hazards, enabling controlled variation in task difficulty and duration. The distributions are predefined and configurable, enabling reproducible experimental conditions across participants. Transition between floors requires coordinated player interaction, enforcing synchronization and shared decision-making before advancing the mission.

To simulate perceptual stress and uncertainty, we incorporate visibility constraints. Object detectability is limited by spatial distance and modulated by player health, reflecting reduced perceptual capability under strain. As health declines, the probability of perceiving nearby objects decreases, introducing uncertainty into search and decision-making. Elevator system and visibility constraints provide structured task control while modeling realistic perceptual challenges.

%% file: figure/04-SAR_features.tex
\begin{figure}[tb]
    \centering
    \includegraphics[width=\columnwidth]{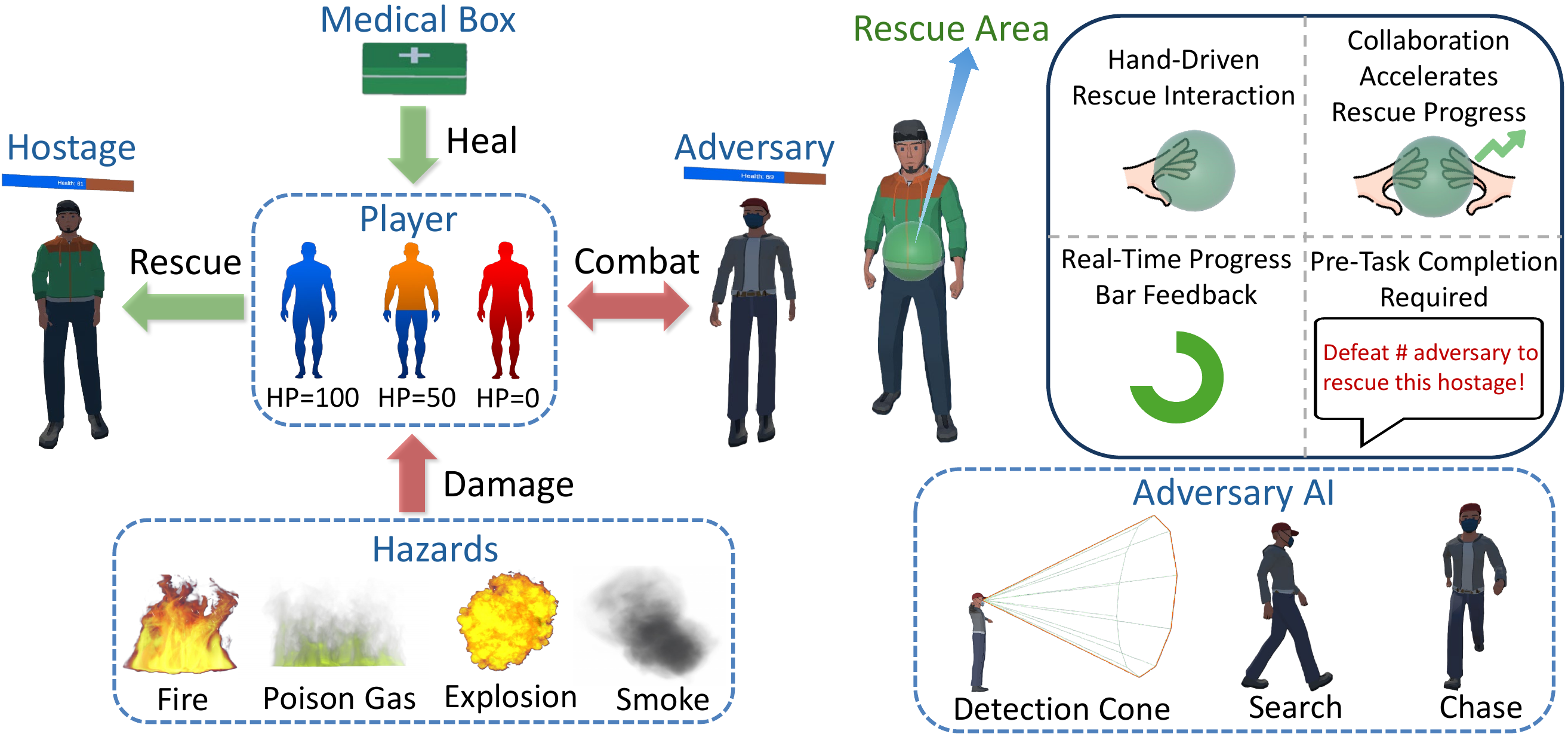}
    \caption{Search-and-rescue features.
    (\textbf{Left}): Unified health system linking hazards, combat, rescue, and recovery.
    (\textbf{Top Right}): Rescue mechanism involving hand-driven interaction, progress feedback, collaboration, and pre-task completion check.
    (\textbf{Bottom Right}): Adversary AI's detect, search, and chase behaviors.}
    \label{fig:SAR_features}
    \vspace{-0.2in}
\end{figure}

%% file: section/06-multiplayer.tex
\section{Multi-User Collaboration}

To enable systematic evaluation of cognitive attacks in realistic operational settings, we extend the \sysname{} testbed to support multi-user MR collaboration. Search and rescue tasks depend on shared situational awareness and coordinated action, which introduce additional synchronization and shared-state dependencies beyond single-user interaction. Our multi-user design provides shared spatial alignment and real-time synchronization, enabling participants to operate within a consistent environment while exposing collaborative attack surfaces such as desynchronization and shared-perception manipulation.

\subsection{Colocation System and Spatial Alignment}

A consistent spatial reference frame is essential for collaborative MR tasks, as misalignment can disrupt shared perception and coordination. To ensure spatial consistency, users first reconstruct their physical environment and join a network session. We then employ shared spatial anchors~\cite{metassa} to align coordinate systems across devices, transforming each local frame into a unified global reference. Once aligned, all participants perceive virtual objects at identical spatial locations relative to the physical environment. Visual feedback confirms successful alignment, ensuring that spawning, animation, and interaction remain consistent across users. This colocation mechanism forms the foundation for reliable shared situational awareness and enables controlled study of collaborative cognitive attack surfaces.

\subsection{Real-Time Synchronization}

To preserve shared situational awareness, the platform synchronizes both virtual objects and player-related states across all participants. Interactive objects, including hostages, adversaries, hazards, and weapons, are instantiated and updated consistently so that all users observe identical environmental dynamics. Concurrent interactions are resolved through authority-controlled state management to prevent divergence across clients. In addition to environmental elements, the system continuously synchronizes player-specific information, including health, six-degree-of-freedom (6DoF) head pose, and mission-level metrics (e.g., adversaries defeated and hostages rescued). These updates are reflected in real time on each participant’s interface, ensuring a consistent perception of team status. This synchronization architecture maintains coherence in the shared environment and is essential for studying collaborative cognitive attack surfaces, where inconsistencies in state propagation, authority control, or perception could directly influence coordination, trust, and decision-making.

%% file: section/07-data.tex
\section{Meta Control and Data Collection Framework}

\subsection{Meta Control}

To facilitate experimental user studies, the \sysname{} testbed integrates meta-level control mechanisms that enable developer testing and centralized coordination.
At the local level, participants interact with a palm-menu interface, while at the global level, a unified server control manages training flow and cognitive attack.

\textbf{Palm Menu Control.}
As \Cref{fig:palm_control} shows, users can summon a palm-mounted menu to switch between modes, manage server connections, and configure study parameters.
The menu provides a training mode that familiarizes participants with virtual objects, interactions, and task mechanics, as well as experiment modes that instantiate predefined scenario configurations (e.g., object types and spatial distributions).
Key parameters (e.g., the number of hostages and adversaries) can be adjusted at runtime, enabling rapid iteration over experimental conditions without rebuilding the scene.

\textbf{Unified Server Control.}
We designed the unified server control to facilitate training and manage cognitive attacks during experiment modes, as \Cref{fig:server_control} shows.
In training mode, the server selectively enables virtual objects and interactions to standardize participant onboarding.
In experiment modes, the server manages mode transitions and centrally coordinates cognitive attack injection, ensuring synchronized delivery across participants and consistent timing across trials.

\input{figure/07-palm}

\subsection{Data Collection Pipeline}
\label{sec:datacollection}

The \sysname{} testbed integrates a synchronized, multimodal data collection pipeline that records system- and user-level metrics in real time, including cognitive attack events, behavioral data, user activity events, and subjective measures.
In addition, the pipeline captures first-person video and in-game audio communication, providing contextual grounding for quantitative analysis.
The server stores all data for subsequent processing, providing a synchronized multimodal dataset that supports fine-grained post-hoc analysis of user resilience, decision-making, and collaboration behaviors under adversarial MR conditions.
The data collection pipeline is shown in~\Cref{fig:data_collection}.

\input{figure/08-pipeline}

\textbf{Real-Time Data Collection.}
During experiment modes, the platform continuously streams system-level and user-level data to the centralized server.
Cognitive attacks are injected at controlled intervals, and each perturbation event is explicitly time-stamped and labeled by type and intensity.
This structured annotation enables direct alignment between attack exposure and subsequent behavioral or subjective responses, supporting systematic evaluation of attack impact.

In parallel, the system captures rich behavioral data, including head and eye tracking.
Specifically, the system continuously records the position and rotation of the head and eyes, gaze direction, and 2D focal points, along with derived metrics such as gaze targets and information about visible virtual objects (e.g., object count, name, and pixel coordinates within the user’s field of view).
These data provide fine-grained insights into users’ attention allocation and cognitive load.

To collect subjective measures such as engagement and trust, synchronized pop-up questionnaires (shown in ~\Cref{fig:questionnaire}) appear in participants’ heads-up display (HUD) at fixed intervals.
Questionnaire presentation is synchronized across all participants to ensure temporally aligned reporting and consistent exposure windows.
Responses use a 7-point Likert scale, with a default value of 0 indicating an invalid response to prevent bypassing.

The system also logs user interaction events in real time.
Actions such as defeating adversaries, rescuing hostages, or using medical boxes are time-stamped and annotated with structured event labels, enabling precise reconstruction of behavioral trajectories and decision sequences during post-hoc analysis.

\input{figure/09-questionnaire}

\textbf{Data Logging and Post-Processing.}
All participant data streams are collected through a centralized Python-based server prior to local storage.
Time-stamps are aligned across behavioral logs, cognitive attack events, questionnaire responses, and first-person video, enabling cross-modal synchronization.
A post-processing pipeline integrates these heterogeneous data sources into a unified dataset structured for quantitative and qualitative analysis.
The modular schema allows additional attack types or behavioral signals to be incorporated without restructuring the logging framework, supporting extensibility for future adversarial MR studies.

%% file: figure/07-palm.tex
\begin{figure}[tb]
    \centering
    \subfloat[Palm menu control.]{\includegraphics[height=0.24\linewidth]{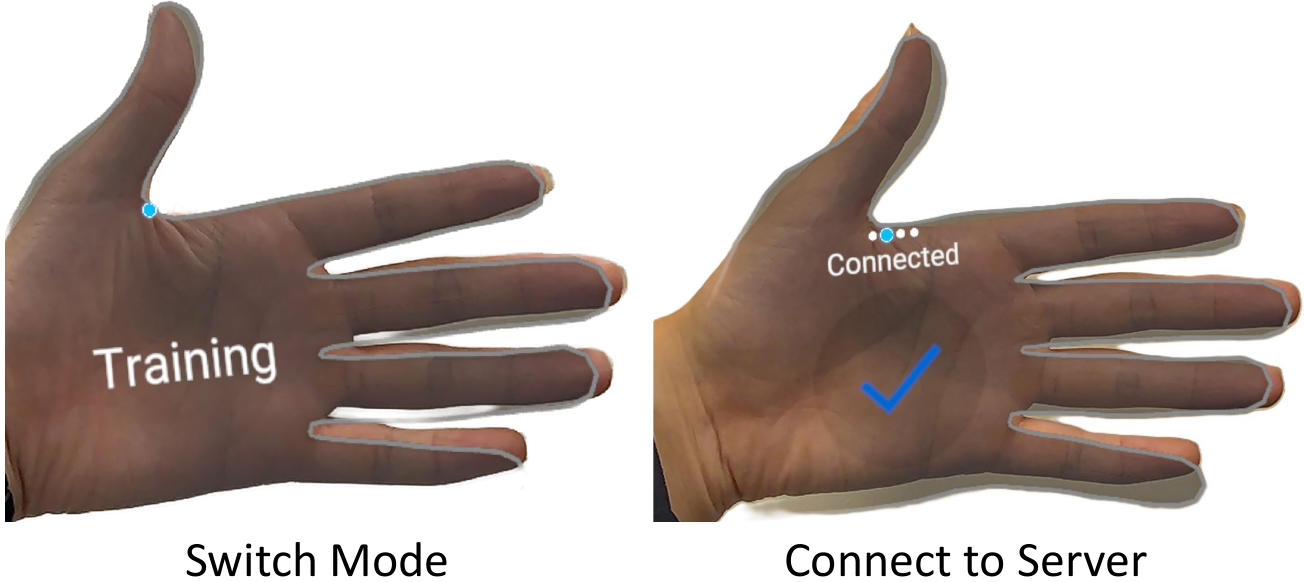}\label{fig:palm_control}}
    \hspace{0.1in}
    \subfloat[Unified server control.]{\includegraphics[height=0.24\linewidth]{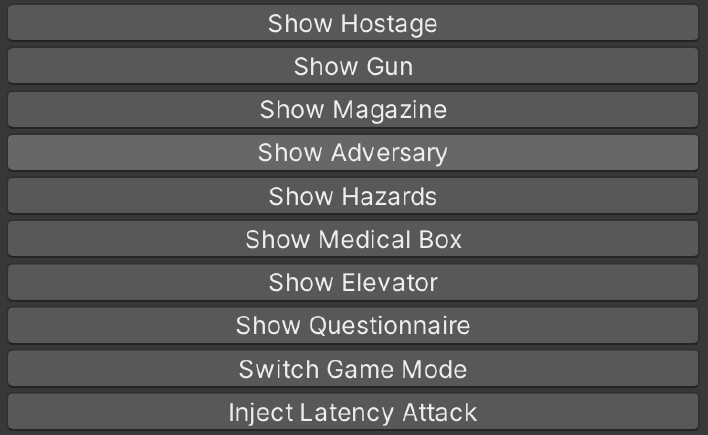}\label{fig:server_control}}
    \caption{Meta control mechanisms.}
    \label{fig:meta_control}
    \vspace{-0.2in}
\end{figure}

%% file: figure/08-pipeline.tex
\begin{figure}[tb]
    \centering
    \includegraphics[width=\columnwidth]{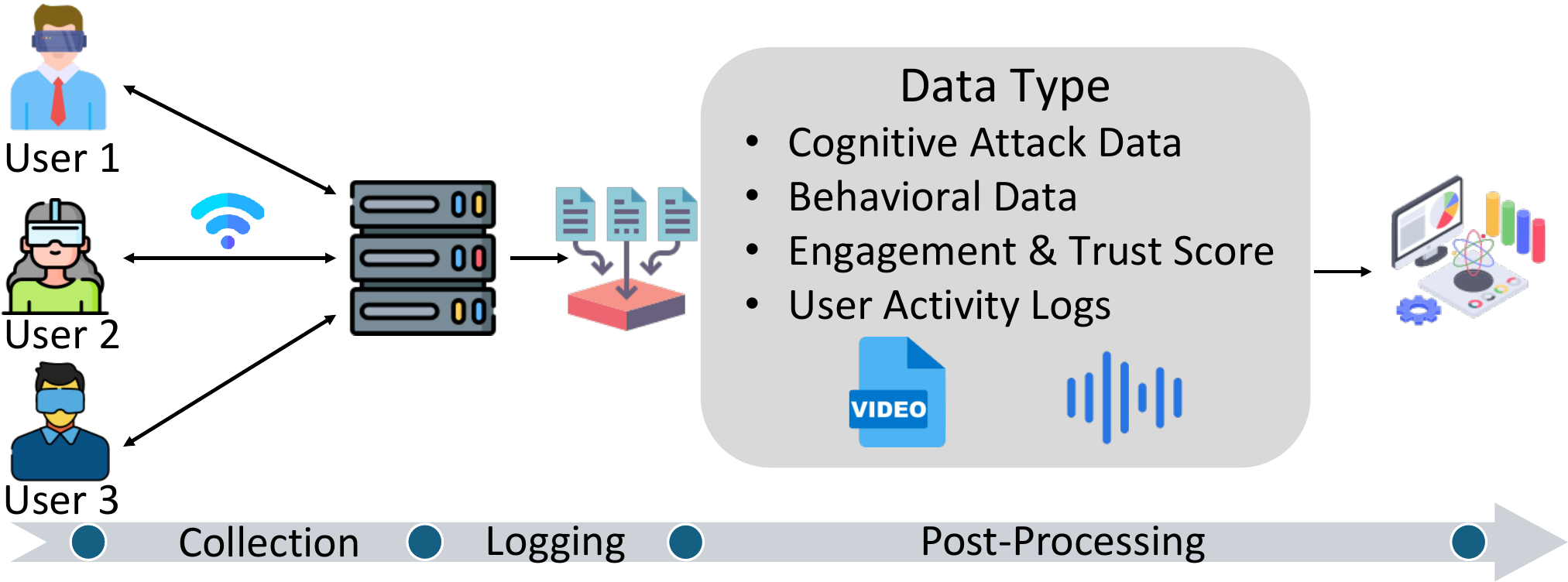}
     \caption{Data collection pipeline. Multimodal data are captured and logged on a centralized server for post-hoc analysis.}
    \label{fig:data_collection}
    \vspace{-0.15in}
\end{figure}

%% file: figure/09-questionnaire.tex
\begin{figure}[tb]
    \centering
    \includegraphics[width=\columnwidth]{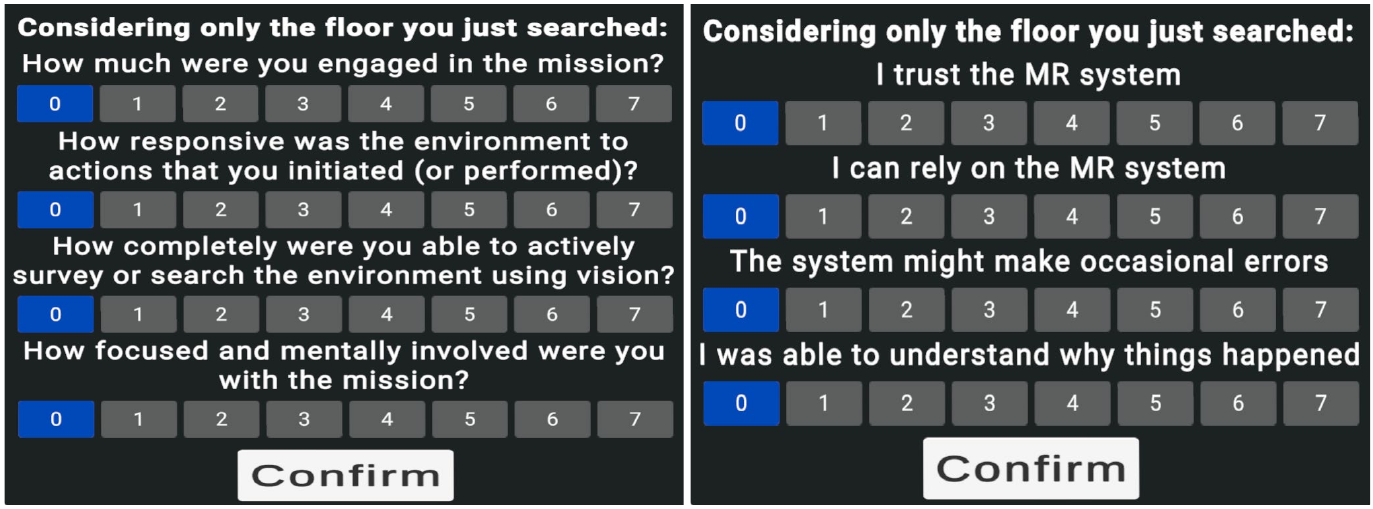}
     \caption{Pop-up questionnaire mechanism. At predetermined intervals, the system displays a pop-up to collect 7-point Likert scale ratings of user engagement and trust.}
    \label{fig:questionnaire}
    \vspace{-0.2in}
\end{figure}

%% file: section/08-user_study.tex
\section{User Study: Latency-Based Cognitive Attack}
While the \sysname{} testbed supports a broad range of cognitive attacks across multiple categories, conducting comprehensive human-subject evaluations for all attack types within a single paper is infeasible. Instead, we focus our user study on latency-based display perturbation, specifically the frame rate reduction and action-to-display latency mechanisms described in~\Cref{sec:cog_attack}. This attack serves as a representative evaluation case within our framework because it directly interferes with the perception-action feedback loop essential for situational awareness and coordination in MR environments.

We conducted a controlled human-subject study with 22 participants to systematically examine the impact of latency manipulation while demonstrating the framework's capability to support controlled human-subject evaluation. The study integrates objective task-performance metrics, fine-grained behavioral signals, and subjective measures of engagement and trust. This multi-level measurement design enables a comprehensive analysis of how display latency affects perception, decision-making, and collaborative performance in MR-based search-and-rescue tasks. The same experimental protocol and data-collection pipeline can be readily extended to evaluate other cognitive attack types supported by \sysname{} in future work.

\subsection{Research Questions for the User Study}\label{sec:user_study_RQ}
The user study is designed to investigate how the latency-based cognitive attack affects user performance, behavior, and subjective experience during collaborative MR search-and-rescue tasks. Specifically, we address the following research questions:
\begin{itemize}[nosep]
    \item \textbf{RQ1:} \emph{How does injected latency affect task performance in collaborative MR search-and-rescue scenarios?}
    \item \textbf{RQ2:} \emph{How do users adapt their behavior under increasing latency during collaborative MR tasks?}
    \item \textbf{RQ3:} \emph{How does latency-based cognitive attack affect users' subjective experience, including engagement and trust?}
\end{itemize}

These research questions aim to provide a comprehensive understanding of the cognitive and behavioral impact of the latency-based attack in MR training environments, while demonstrating the capability of the \sysname{} testbed to support systematic, human-centered evaluation of cognitive attacks.

\subsection{Participants}
We recruited 22 participants for the user study (6 female, 15 male, 1 non-binary), ranging in age from 19 to 32 years, with an average age of 24.4 and a standard deviation of 3.8.
Participants had normal or corrected-to-normal vision and no self-reported motor impairments that would interfere with MR interaction. All participants provided informed consent prior to participation, and the study protocol was approved by the local Institutional Review Board (IRB).

Participants completed the study as part of two-person teams in a collaborative MR search-and-rescue session, labeled as Participant A and Participant B for analysis. During the task, teammates were required to coordinate navigation, adversary engagement, and hostage rescue under varying latency conditions. Each participant followed the same structured protocol consisting of onboarding, a training phase, and experimental trials conducted under controlled conditions to reduce variability due to learning effects. Participants received \$40 cash compensation for their time.

\subsection{Experimental Design}
\input{figure/12-timeline}
\textbf{Study Overview.}
The user study followed a structured experimental design lasting approximately 120 minutes per participant. The session consisted of a preparation and training phase, two collaborative search-and-rescue missions, two intervening breaks, and a post-study feedback phase, as shown in~\Cref{fig:timeline}. Participants were additionally invited to complete an optional online follow-up survey the following day to report any perceived after-effects. This design balances realistic task dynamics with controlled latency manipulation while mitigating fatigue in a long-duration MR study.

\textbf{Preparation and Training.}
Participants first completed a standardized preparation and training phase designed to familiarize them with the MR hardware, interaction techniques, and task mechanics, thereby reducing learning effects during the formal study. Training covered core interactions required during the study, including hand-based interaction, adversary engagement, hostage rescue, hazard avoidance, and use of the virtual elevator system. Participants also practiced responding to pop-up engagement and trust questionnaires presented within the MR environment. The training protocol followed a fixed script and was identical across sessions to ensure procedural consistency.

\textbf{Search-and-Rescue Missions.}
Participants completed two collaborative search-and-rescue missions, each lasting approximately 25 minutes, within a virtual five-floor building. During each mission, teams searched for and rescued hostages while managing hazards and adversaries under time pressure. Each virtual floor lasted approximately four minutes. Upon completion of each floor, participants responded to synchronized pop-up questionnaires assessing engagement and trust, resulting in five prompts per mission.

To ensure fair comparisons across latency conditions and participants, a predefined mission configuration was assigned to each floor.
While the spatial locations of virtual objects varied across floors to reduce memorization effects, the number and types of virtual objects remained consistent to preserve comparable task composition.
Each floor corresponded to a distinct latency condition.
Latency-based cognitive attack was applied by perturbing display responsiveness, including controlled action-to-display delays and reduced frame rates, enabling systematic analysis of latency effects on task performance, behavior, and subjective experience.

\textbf{Breaks and Feedback.}
Two 10-minute breaks were scheduled between missions, during which participants removed the MR headset and completed questionnaires. 
These breaks were intended to reduce physical and cognitive fatigue and mitigate potential simulator sickness.
Following the second break, participants completed demographic questionnaires and provided general feedback on their experience.

\textbf{Follow-on Survey.}
The day after the study, participants were invited to complete a brief optional online survey evaluating any after-effects within 12 hours following participation.
The survey consisted of a binarized form of the Simulator Sickness Questionnaire (SSQ)~\cite{kennedy1993simulator}.

\subsection{Measures}
To evaluate the impact of latency-based cognitive attack on collaborative MR performance and user experience, we analyze objective task performance metrics, behavioral indicators, and subjective measures collected during the study.

\textbf{Objective Task Performance.}
Objective task performance was evaluated for each search-and-rescue mission, including:
(1) \emph{Mission completion outcomes}, including the team/floor-level hostage rescue rate and the number of adversaries defeated; (2) \emph{Task efficiency}, measured by the team/floor-level time to first hostage rescue, and the exploratory rescue-event-level per-rescue completion time; (3) \emph{Safety-efficiency trade-offs}, quantified through hostage health loss events relative to rescue timing.
These metrics reflect how latency perturbations affect participants’ ability to complete mission objectives under time pressure.

\textbf{Behavioral Measures.}
Behavioral indicators were derived from in-mission activity, including: (1) \emph{Interaction behavior}, such as the frequency of weapon-use actions; (2) \emph{Exposure and state-related behavior}, assessed through hazard exposure events.
These behavioral measures characterize how participants adapt their interaction strategies under varying levels of latency.

\textbf{Subjective Measures.}
Participants’ subjective experience was evaluated using self-reported engagement and trust ratings collected through repeated in-task questionnaires and post-study surveys:
\emph{Engagement and trust}, measured as participant/floor-level repeated measures through pop-up questionnaires after each mission floor, yielding five responses per mission.
These responses capture both momentary fluctuations and cumulative effects of latency-based cognitive attacks on user experience.

\subsection{Results}
We report results from the 22-participant user study examining the effects of the latency-based cognitive attack on collaborative MR performance, behavior, and subjective experience. Although two missions were scheduled, some teams did not complete the second mission within the session time. Therefore, all reported analyses are based on data from the first mission to ensure consistency across participants. In this mission, latency levels were assigned to floors in a fixed sequence (0, 17, 62, 9, and 25 ms from floors 1 to 5). Results are organized according to task performance, behavioral adaptation, and subjective measures, corresponding to the research questions outlined in~\Cref{sec:user_study_RQ}.

\textbf{Task Performance.}
\input{figure/13-task_perfomance}
\Cref{fig:taskPerformance} summarizes team-level task performance under different injected latency levels. Overall, the results suggest that participants adapted their task strategies under higher latency, allowing teams to maintain high rescue rates despite increased perceptual and interaction delays. However, latency continued to influence task dynamics and rescue quality, revealing a trade-off between successful task completion and the quality of task outcomes.

\emph{Rescue Rate.}
As shown in~\Cref{fig:taskPerformance} (Left), rescue rate differs significantly across latency conditions (Friedman test: $\chi^2(4)=30.07$, $p<0.001$, Kendall's $W=0.68$; generalized estimating equations (GEE) Wald test: $\chi^2(4)=45.28$, $p<0.001$). Interestingly, the highest rescue rates are observed under the strongest latency perturbations (25 ms and 62 ms), a counter-intuitive result given that increased latency would generally be expected to impair task performance.
As discussed in the following analysis of time-to-first-rescue, participants adapted their task strategies under higher latency by initiating rescues earlier in the mission, which likely compensated for increased perceptual and interaction delays and contributed to the observed rescue performance.

\emph{Adversaries Defeated.}
Although the total number of adversaries defeated is evaluated as a team/floor-level metric, only one weapon is available per floor.
Therefore, in addition to the team/floor-level metric, \Cref{fig:taskPerformance} (Center left) presents the distribution of adversaries defeated at the participant level to illustrate how combat responsibility is allocated between teammates under different latency conditions.
The median number of adversaries defeated remains relatively stable across conditions, with no clear monotonic relationship between latency and adversaries defeated. However, higher latency levels (particularly 25 ms) exhibit increased dispersion for certain participants, indicating greater variability in individual combat engagement. This pattern suggests that latency does not uniformly reduce adversary defeat rates, but may influence how participants allocate effort between combat and rescue activities.

\emph{Rescue Time.}
As shown in~\Cref{fig:taskPerformance} (Center right), the time to first rescue differs significantly across latency conditions (Friedman test: $\chi^2(4)=17.00$, $p=0.0019$, Kendall's $W=0.53$). A linear mixed-effects model also identified significantly shorter times to first rescue under the 9 ms ($p=0.001$), 25 ms ($p<0.001$), and 62 ms ($p<0.001$) latency conditions relative to the baseline. Interestingly, teams generally initiated rescues earlier under higher latency conditions rather than exhibiting progressively slower performance. These findings help explain the rescue-rate results discussed above. By prioritizing rescue objectives earlier in the mission, participants were able to compensate for perceptual and interaction delays, resulting in consistently high, and in some cases higher, rescue rates despite stronger latency perturbations. A similar trend is observed in the exploratory team-level per-rescue completion time analysis (see~\Cref{tab:rescue_time}), where the 25 ms and 62 ms conditions exhibit lower average completion times than several lower-latency conditions. These results suggest that latency primarily influenced task strategy rather than simply degrading task execution.

\input{table/2-rescue_time}

\emph{Rescue Time and Hostage Health Trade-off.}
Although participants adapted their rescue strategies under higher latency conditions, longer rescue durations remained associated with reduced rescue quality.
As shown in~\Cref{fig:taskPerformance} (Right), rescue duration is negatively correlated with hostage health at the time of rescue ($r = -0.28$, $p \approx 0.0042$),
highlighting a time-critical trade-off inherent to the task. This relationship persists across latency conditions, suggesting that while participants adapt to latency and maintain rescue rate, longer rescue durations continued to be associated with lower remaining hostage health.
Overall, these findings indicate that maintaining rescue rate does not eliminate the quality cost associated with longer rescue durations.

\textbf{Behavioral Adaptation.}
\input{figure/14-behavior}
\Cref{fig:behavioralEffects} shows how participant behavior changes under different latency levels, focusing on weapon interaction patterns and exposure to environmental hazards. Unlike task performance metrics, these behavioral measures reveal behavioral adaptations in response to latency-based cognitive attacks.

\emph{Weapon Usage Patterns.}
As shown in~\Cref{fig:weapon_usage_latency}, weapon actions exhibit a non-linear relationship with latency. Moderate latency levels (e.g., 17 and 25 ms) are associated with an increase in both shots fired and reload actions, suggesting increased action frequency under moderate latency. This behavior may reflect perceived uncertainty in action timing and feedback, leading users to fire more frequently or reload proactively to maintain combat readiness. At the highest latency level (62 ms), however, total weapon actions decrease, consistent with a more conservative pattern of combat engagement. These trends suggest that latency does not merely slow interactions but may also influence how participants regulate engagement intensity during combat.

\emph{Hazard Exposure Time.}
~\Cref{fig:hazard_latency} shows the distribution of hazard exposure time across latency conditions, capturing the duration participants spend in fire or poison states. The no-latency condition exhibits the greatest spread in exposure duration, whereas moderate latency levels (17 ms and 25 ms) show lower median exposure times and comparatively tighter distributions. The highest latency level (62 ms) demonstrates moderate variability but does not exceed that observed in the no-latency condition. This pattern suggests more consistent hazard avoidance at moderate latency levels, while larger delays appear to affect responsiveness during hazard avoidance. These findings indicate that latency may influence the consistency of hazard avoidance behavior rather than producing systematic degradation.

\textbf{Subjective Experience.}
\input{figure/15-subjective}
\Cref{fig:subjectiveEffects} presents participants’ subjective responses under different latency conditions, focusing on perceived engagement and trust in the system. In contrast to task performance and behavioral measures, these subjective ratings capture participants’ perceived interaction quality and system reliability.
Despite statistically significant differences across latency conditions, engagement and trust remain generally stable, suggesting a limited impact on participants' subjective experience.

\emph{Engagement.}
As shown in~\Cref{fig:engagement_latency}, engagement differs significantly across latency conditions (Friedman test: $\chi^2(4)=29.37$, $p<0.001$, Kendall's $W=0.33$). A linear mixed-effects model similarly identified significantly higher engagement scores under all latency conditions relative to the baseline (17 ms: $p=0.013$; 9 ms, 25 ms, and 62 ms: all $p<0.001$). Interestingly, participants reported slightly higher engagement under latency than under the no-latency condition, despite the added perceptual and interaction challenges. Although these differences are statistically significant, average engagement scores remained consistently high across all conditions. These findings suggest that the evaluated latency-based cognitive attacks did not reduce perceived engagement and may have slightly increased task involvement.

\emph{Trust.}
As shown in~\Cref{fig:trust_latency}, trust differs significantly across latency conditions (Friedman test: $\chi^2(4)=11.51$, $p=0.021$, Kendall's $W=0.13$). A linear mixed-effects model similarly identified slightly higher trust scores under the 9 ms ($p=0.004$), 25 ms ($p<0.001$), and 62 ms ($p=0.002$) latency conditions relative to the baseline, while the 17 ms condition did not differ significantly ($p=0.146$). Despite these statistically significant differences, trust scores remained relatively stable across latency conditions, with similar median values and substantial overlap in their distributions. These findings suggest that the evaluated latency-based cognitive attacks did not substantially reduce perceived trust, and the observed differences were relatively modest.

\emph{Self-reported After-effects}.
10 out of 22 participants completed the optional binarized SSQ the day after the study. Only one participant reported a headache within 12 hours of participation. Given the limited responses, further investigation is needed to determine whether \sysname{} may contribute to post-session symptoms.

%% file: figure/12-timeline.tex
\begin{figure}[tb]
    \centering
    \includegraphics[width=\columnwidth]{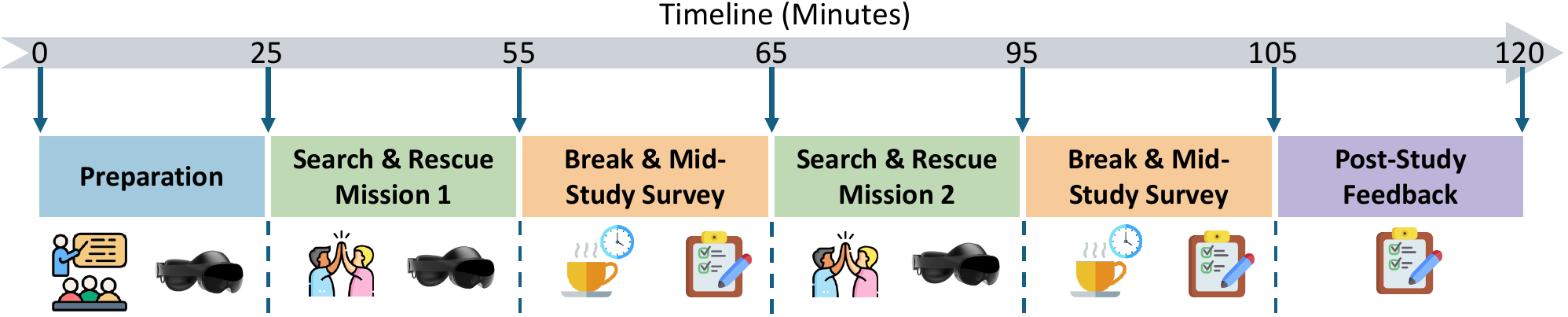}
    \caption{User study timeline.}
    \label{fig:timeline}
    \vspace{-0.2in}
\end{figure}

%% file: figure/13-task_perfomance.tex
\begin{figure*}[tb]
\centering
\includegraphics[height=0.175\textwidth]{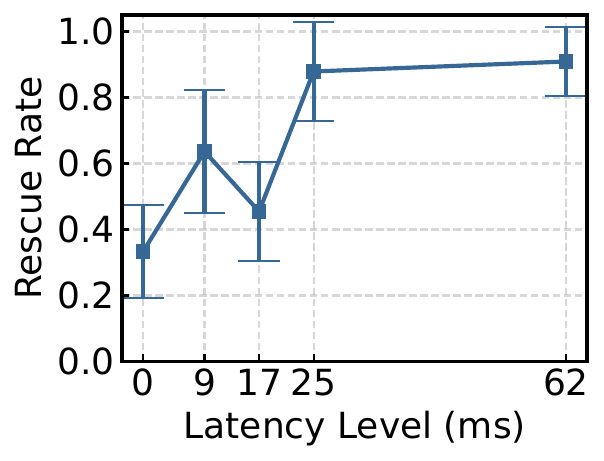}\label{fig:rescue_rate_latency}
\hfill
\includegraphics[height=0.175\textwidth]{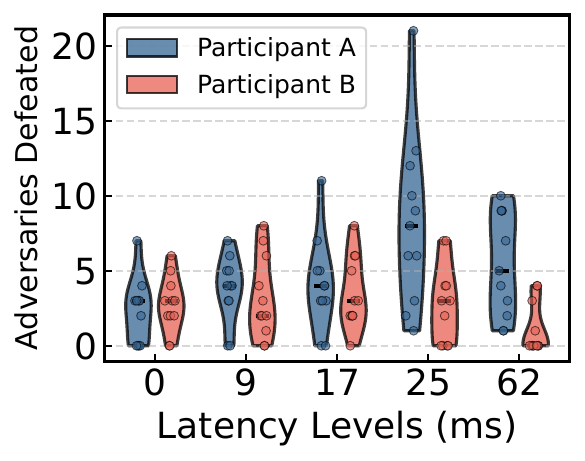}\label{fig:adversaries_latency}
\hfill
\includegraphics[height=0.175\textwidth]{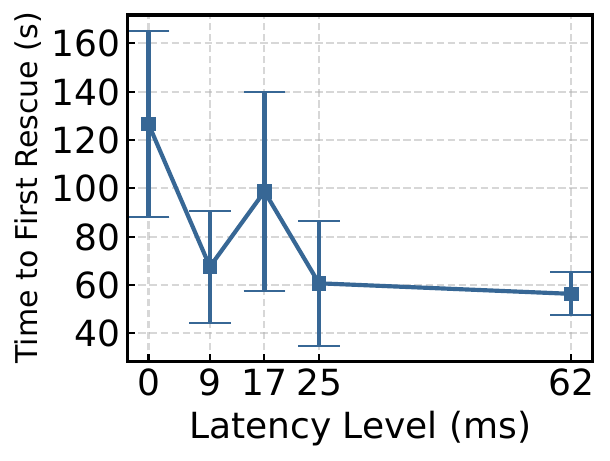}\label{fig:time_first_rescue}
\hfill
\includegraphics[height=0.175\textwidth]{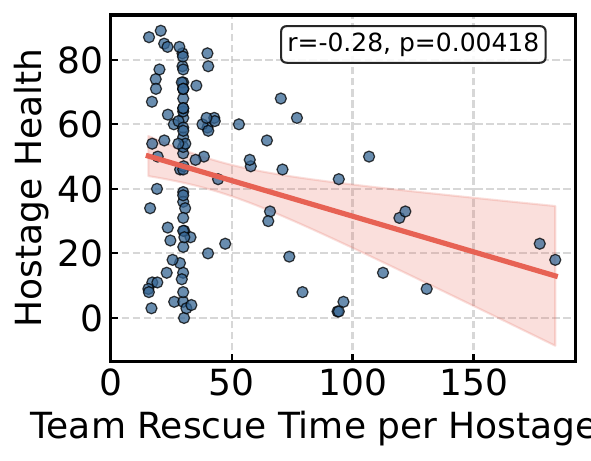}\label{fig:rescue_hhp}

\caption{Task performance under latency-based cognitive attack.
(\textbf{Left}): Team rescue rate as a function of injected latency, showing mean and 95\% confidence intervals.
(\textbf{Center left}): Distribution of adversaries defeated per participant under different latency levels.
(\textbf{Center right}): Time to first successful rescue across latency conditions.
(\textbf{Right}): Relationship between team-level rescue duration and hostage health at rescue, with linear regression and 95\% confidence band.
}
\label{fig:taskPerformance}
\vspace{-0.35in}
\end{figure*}

%% file: table/2-rescue_time.tex
\begin{table}[t]
\centering
\caption{Team-Level Rescue Time per Hostage by Latency Level}
\label{tab:rescue_time}
\begin{tabular}{c c c c c}
\hline
\textbf{Latency (ms)} & \textbf{Count} & \textbf{Mean (s)} & \textbf{SD (s)} & \textbf{Median (s)} \\
\hline
0  & 11 & 56.152 & 41.276 & 33.034 \\
9  & 21 & 40.113 & 28.192 & 29.999 \\
17 & 15 & 61.760 & 52.994 & 37.952 \\
25 & 28 & 34.571 & 19.572 & 30.007 \\
62 & 30 & 36.957 & 20.283 & 30.025 \\
\hline
\vspace{-0.3in}
\end{tabular}
\end{table}

%% file: figure/14-behavior.tex
\begin{figure}[tb]
    \centering
    \subfloat[Weapon usage under latency.]{%
        \includegraphics[width=0.45\linewidth]{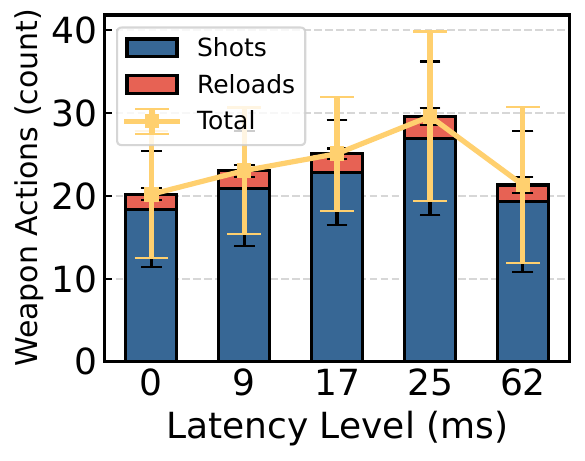}%
        \label{fig:weapon_usage_latency}
    }
    \hspace{0.1in}
    \subfloat[Hazard exposure under latency.]{%
        \includegraphics[width=0.45\linewidth]{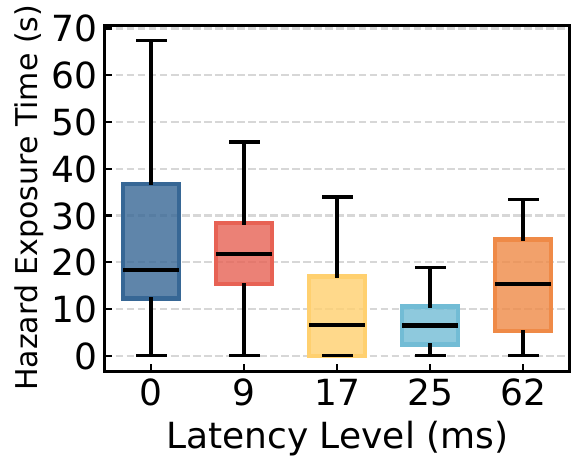}%
        \label{fig:hazard_latency}
    }
    \caption{Behavioral adaptation under latency-based cognitive attack.
    (\textbf{Left}): Changes in weapon interaction patterns, including shots fired and reload actions. 
    (\textbf{Right}): Distribution of hazard exposure time (fire and poison) across latency conditions.
    }
    \label{fig:behavioralEffects}
    \vspace{-0.3in}
\end{figure}

%% file: figure/15-subjective.tex
\begin{figure}[tb]
    \centering
    \subfloat[Engagement vs. latency.]{%
        \includegraphics[height=0.35\linewidth]{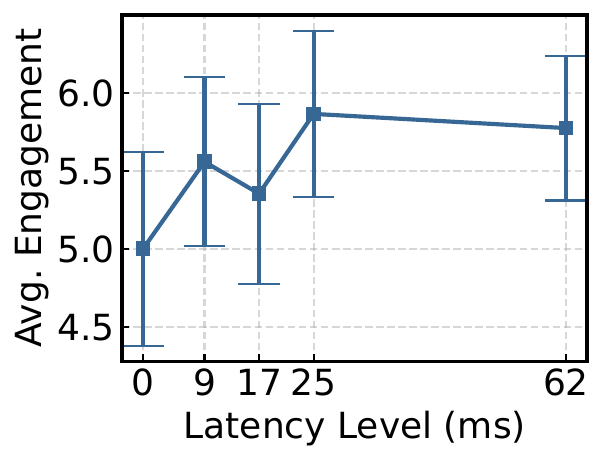}%
        \label{fig:engagement_latency}
    }
    \hspace{0.1in}
    \subfloat[Trust vs. latency.]{%
        \includegraphics[height=0.35\linewidth]{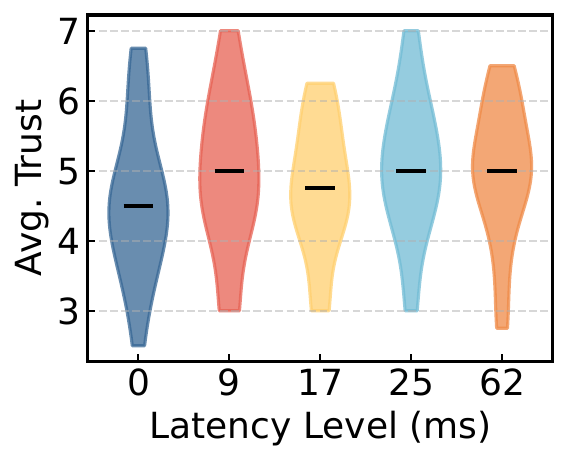}%
        \label{fig:trust_latency}
    }
    \caption{Subjective experience under latency-based cognitive attack.
    (\textbf{Left}): Mean engagement scores across latency conditions.
    (\textbf{Right}): Distribution of trust scores across latency conditions.
    }
    \label{fig:subjectiveEffects}
    \vspace{-0.3in}
\end{figure}

%% file: section/09-discussion.tex
\section{Discussion and Future Work}
\label{sec:discussion}

This section discusses current limitations of the framework and outlines directions for future research and system enhancement.

\textbf{Experimental Design Considerations.} The current user study employs a fixed ordering of latency conditions to ensure identical task configurations across participants. While this design introduces the possibility of order effects, the observed results are unlikely to be explained solely by participant familiarization. Repeated-measures analyses identified significant effects of latency on both rescue rate and time to first rescue, and the behavioral analyses suggest that participants adapted their task strategies under higher latency conditions by initiating rescues earlier. Randomized condition ordering would further reduce potential order effects and will be considered in future studies, although such a design would require substantially more participants to maintain balanced experimental conditions and statistical power.

\textbf{Attack Detection and Mitigation.}
Our proposed testbed provides a unified platform for injecting multiple cognitive attacks within a controlled MR environment, enabling system-level evaluation under adversarial conditions.
However, the current implementation focuses on systematic attack deployment and evaluation rather than real-time detection or mitigation.
Although detection modules are not yet integrated, the modular architecture of the testbed is explicitly designed to support such extensions. Future work will incorporate detection and response mechanisms to enable closed-loop evaluation of adversarial robustness.
For example, prior works~\cite{xiu2025detecting, xiu2024lobstar, xiu2025viddar} have explored visual-language model (VLM)-based approaches for detecting adversarial or misleading AR content, which could be integrated as detection components and systematically evaluated within our platform.

\textbf{Scalability and Generalization.}
Although we use collaborative search-and-rescue as a representative application scenario, the proposed framework is not limited to this domain. Search-and-rescue was selected because it integrates multiple core components of collaborative MR applications, including target search and localization, navigation under hazards, combat interaction, multi-user collaboration, synchronization, decision-making, and behavioral adaptation. Furthermore, the framework adopts a modular architecture that separates attack injection, synchronization and colocation, experiment orchestration, data logging, and task content, enabling adaptation to other collaborative MR applications such as industrial target search, fire evacuation training, and tactical shooting training. In future work, generative AI techniques~\cite{11339452} could be leveraged to accelerate scenario construction and virtual content generation, reducing manual design effort and enabling more diverse task configurations. While the current user study focuses on latency-based perturbations as a representative case study, future work will evaluate the remaining cognitive attack classes implemented in the framework to systematically investigate how diverse adversarial conditions influence users' perception, behavior, coordination, and trust in collaborative MR settings.

\textbf{Multimodal Behavioral Analysis.}
The testbed supports synchronized collection of multimodal data, enabling comprehensive analysis of user behavior during collaborative MR tasks. This extensible data collection pipeline facilitates quantitative evaluation of both task performance and user experience under adversarial or degraded conditions rather than in isolated settings. Furthermore, these multimodal data can support future research on how cognitive attacks influence visual attention, situational awareness, and decision-making in collaborative MR environments.

%% file: section/10-conclusion.tex
\section{Conclusion}
We presented the \sysname{} testbed, an end-to-end multi-user mixed reality framework for systematic evaluation of cognitive attacks in collaborative search-and-rescue training. The framework adopts a modular architecture that supports synchronized collaboration, integrates four representative attack classes, and incorporates a multimodal data collection pipeline for analyzing user behavior and performance under adversarial conditions. Through a controlled user study of the latency-based attack, we demonstrated measurable effects on task performance and user behavior while characterizing users' subjective experience under adversarial conditions.
Looking forward, \sysname{} provides a reproducible foundation for collaborative MR security research by enabling controlled attack injection, synchronized behavioral measurement, and systematic evaluation of adversarial effects.

%% file: SupplementaryMaterial.tex
\section{Implementation Details}

\textbf{Hardware.}
The system runs on Meta Quest 3 and Meta Quest Pro headsets (for eye tracking). 
The server/control machine is a Dell G15 laptop running Python 3.9.

\textbf{Software.}
The application is developed in Unity 2021.3.33f1 (LTS).
We use Meta XR Core SDK, Meta XR All-in-One SDK (v67.0.0), Meta Mixed Reality Utility Kit (MRUK, v67.0.0), Meta Shared Spatial Anchors, Meta Passthrough API, and Meta Eye Gaze API (Quest Pro only).
Multi-user synchronization is implemented using Photon Fusion 2.0.3.

\textbf{System Architecture.}
The client (HMD) handles spatial mapping, rendering, interaction, and local sensing (head/hand pose and eye gaze when available).
Photon Fusion synchronizes virtual objects and player states across users via authority-controlled state management.
The server manages experiment control and log storage for post-hoc analysis.

\section{Search-and-Rescue Features Details}

\subsection{Health System}

A modular health point (HP) framework shown in \Cref{fig:health} manages the status of players, hostages, and adversaries, providing a consistent mechanism for survival and interaction.
The system combines passive decay with event-driven health changes to create engaging dynamics in the search and rescue scenario.
Players experience gradual health decay over time, as well as health loss from environmental hazards such as fire, poison gas, and explosions.
In addition, combat interactions with adversaries directly inflict damage, making the player’s HP sensitive to both the environment and active encounters.
Players can restore health only through medical boxes, which serve as critical recovery resources.
Hostages also lose HP over time, and their survival duration varies across individuals.
This variability requires players to prioritize search and rescue efforts before a hostage’s health is depleted.
Adversaries take damage through combat with the player, with their health influenced either by weapon attacks or by simulated fighting mechanics, which \Cref{sec:adversary_combat} discussed.

The user receives visual feedback through multiple health indicators.
Players monitor their own status using a human-shaped health indicator and a numerical display in the head-up display (HUD).
The health of hostages and adversaries is visible through health bars with numerical values positioned above their heads.
All health indicators employ color transitions from blue to orange to signal declining health, with the player’s indicator turning red upon reaching zero HP.
For hostages, the health bar also displays terminal states, switching to ``RESCUED'' when successfully saved or ``DEAD'' when their health is fully depleted.
When a player’s HP reaches zero, they lose the ability to interact with most virtual objects, such as rescuing hostages or using weapons, and must search for a medical box to recover health and regain functionality.

\input{figure/04-HP}

\subsection{Hostile Environment}

As shown in \Cref{fig:health}, we simulate a hostile environment through four types of hazards: fire, poison gas, explosion, and smoke.
The fire rendering has dynamic flames and sparks for realism, and it causes continuous damage when a player enters its radius.
An ``On Fire'' indicator in the HUD provides immediate feedback when damage occurs.
Fire also has a probabilistic spreading mechanism to nearby positions, simulating the stochastic nature of real fire propagation.
Poison gas causes damage similar to fire but introduces residual effects that continue to reduce health even after a player leaves the gas zone.
A ``Poisoned'' indicator in the HUD alerts the player when toxic effects are active.
Explosions occur at fire locations with an adjustable probability, producing substantial area damage within their blast radius.
Smoke blocks vision and prevents the detection of virtual objects such as hostages, adversaries, weapons, and medical boxes, as well as obscuring the physical environment.
This simulates severely degraded conditions for search and rescue.
These hazards collectively establish a hostile and unpredictable environment, which intensifies both the challenge and immersion of the tactical scenario.

\input{figure/05-NPC}

\subsection{Rescue Mechanism}

As shown in~\Cref{fig:hostage_animation} (Left), hostages have animation transitions through a sequence of states, including standing, terrified, kneeling, waving, rescued, and dead.
We designed these animations to transition smoothly.
For example, a standing hostage cannot directly switch to kneeling, since intermediate poses (e.g., terrified) are required to preserve continuity.
When a player appears in front of a hostage, the animation transitions to waving as a signal for rescue.
Since hostages have health, they will eventually enter the dead state if players fail to rescue them in time.
Conversely, a successful rescue triggers the rescued animation, where the hostage performs a thanking gesture toward the player.

\Cref{fig:rescue_mechanism} (Center) illustrates the rescue mechanism.
We realized this process as a hand-driven interaction within a designated rescue zone, represented by a green sphere in front of the hostage.
When a user places a hand into this zone, a progress bar provides real-time feedback.
Rescue is successful once the bar is filled, while withdrawing the hand causes the progress to decrease.
To encourage collaborative behavior, the system increases progress at a faster rate when multiple players interact simultaneously, whereas a single player using both hands does not provide any additional benefit.

To involve interaction with adversaries and increase mission difficulty, a pre-task completion check requires players to defeat a specified number of adversaries before rescuing a hostage.
If this condition is unmet, a warning message appears in the HUD stating ``Defeat \# adversary to rescue this hostage!''
The required number of adversaries increases with each successful rescue, ensuring that the challenge scales as the mission progresses.

\subsection{Adversary AI and Combat System}
\label{sec:adversary_combat}

As \Cref{fig:adversary} (Right) illustrates, the adversary performs a range of intelligent behaviors within the MR environment.
A finite-state machine manages a combination of adaptive behaviors and reactive animations, enabling adversaries to adapt dynamically to player actions and maintain immersive, context-sensitive interactions.
In the absence of players, the adversary wanders randomly to simulate search behavior.
A cone-shaped field of view defines its detection zone.
When a player enters this zone, the adversary transitions to a chase state, running toward the player and causing damage within close range.
In addition to search and chase behaviors, the adversary reacts to player attacks.
When hit by bullets, it displays context-specific animations depending on the body region impacted, including left, right, and rear hit reactions.
Upon defeat, the adversary plays a dying animation to indicate elimination.

\input{figure/06-weapon+}

The combat system integrates both a weapon system and simulated fighting mechanics, enabling diverse engagement dynamics.
The weapon system consists of hand-interactive guns and reloadable magazines.
The gun automatically adjusts to align with the player’s hand when grabbed.
Players can then interact naturally by pulling the trigger with their index finger, with each trigger pull firing a laser beam bullet.
The HUD shows the remaining number of bullets, and when out of ammunition, a warning message ``Reload the Gun'' appears.
At this point, players must search for and interact with a magazine to reload.
Bullets cause variable damage depending on the impact region on the body.
Headshots inflict higher damage compared to body hits, as illustrated in~\Cref{fig:weapon} (Left).
This weapon system enhances interactivity and immersion by simulating realistic firearm use.
In addition to weapons, the combat system incorporates simulated fighting mechanics.
When players and adversaries enter close range, they can inflict damage on each other.
All damage parameters are configurable, allowing adjustment based on experimental needs.
This combination of shooting and close-range fighting creates a comprehensive combat experience that balances realism, interactivity, and challenge.

\subsection{Elevator System and Visibility Mechanics}

To expand physical study spaces, we introduce a virtual elevator system that emphasizes the search property and better aligns with the search and rescue scenario. 
The elevator structures the environment into multiple virtual floors, each with handcrafted, distinct distributions of virtual objects such as hostages, adversaries, and hazards. 
This design enriches the virtual scenario with greater diversity and unpredictability. 
while also enabling data collection across tasks of varying durations. Users decide when to transition to the next floor by activating the elevator through a coordinated interaction in a designated control area represented by a blue sphere (see~\Cref{fig:weapon} (Top right)).
All players must place their hands inside the sphere, which displays a shared progress bar similar to the hostage rescue process.
Once the bar is filled, the elevator transports all players to the next virtual floor.

To further simulate perceptual stress and uncertainty, we introduce visibility mechanics.
A simple example is shown in~\Cref{fig:weapon} (Bottom right), where players can only perceive virtual objects within a 3m radius; objects beyond this distance remain hidden.
In addition to distance-based rules, the system also employs a probabilistic model.
The probability of perceiving objects decreases as player HP declines, reflecting reduced perceptual capability under physical strain.
Objects outside the range remain hidden, while those within view appear probabilistically based on health status.
These visibility mechanics introduce uncertainty into detection and decision-making, thereby mimicking real-world perceptual challenges.

\section{Multi-User Collaboration Details}
\subsection{Colocation System and Spatial Alignment}

A robust colocation system ensures that multiple users maintain a consistent spatial understanding of the MR environment, which is critical for seamless collaboration and immersion.
To achieve this, each user first sets up a virtual scene on their Meta headset that matches the physical environment.
The system leverages Meta’s Mixed Reality Utility Kit (MRUK) to access the constructed virtual room and perform scene understanding, enabling awareness of real-world objects such as floor, walls, and tables.
Users are connected to the same network session through the Auto Matchmaking building block in the Meta XR Core SDK, which automatically joins them to an existing multiplayer room or creates a new one if none is available.
In our implementation, this functionality relies on the shared mode of the Photon Fusion networking framework, which ensures that all users with the same AppID are connected to a common session.
Once networked, the system employs the Shared Spatial Anchors (SSA) API to align coordinate systems across devices.
One headset creates a spatial anchor (e.g., a floor reference) and shares it over the network, while others resolve this anchor to establish a common world origin, thereby transforming all local coordinate systems into the same global frame.
To provide immediate confirmation of successful alignment, the system displays a spatial anchor, represented by coordinate axes, that serves as visual feedback of colocation accuracy.
This colocation pipeline establishes the foundation for consistent spawning, animation, and interaction of virtual objects within the shared environment, which supports seamless multi-user collaboration.

\subsection{Real-Time Synchronization}

The platform synchronizes both virtual objects and player-related states in real time across all users.
The system consistently manages virtual objects such as hostages, adversaries, hazards, and weapons, including their spawning, despawning, animations, and interactions.
At the same time, to maintain a coherent shared environment, the platform continuously updates player-specific data such as health, pose, and hand movements, along with mission-level metrics including the number of hostages rescued and adversaries defeated.
We implemented this synchronization via the Photon Fusion networking framework, where the first client in a session initializes the state and subsequent clients join through peer-to-peer synchronization.
In the following, we describe these two aspects in detail: synchronization of virtual objects and synchronization of player-related states and mission metrics.

\textbf{Virtual objects.}
In our system, all spawnable virtual objects are prefabs with a \textit{NetworkObject} component.
The StateAuthority spawns these objects through the \textit{NetworkRunner}, and Photon Fusion replicates the spawn, including the prefab reference and initial state, so that each client instantiates an identical object.
For objects with animations, we attach the \textit{NetworkMecanimAnimator} component to synchronize their animation states, ensuring that users observe identical behaviors.

By design, the first client is responsible for spawning all virtual objects, while subsequent users synchronize to these states.
This architecture introduces challenges related to StateAuthority.
For instance, if the first client spawns a medical box and holds StateAuthority over it, another user who attempts to consume the medical box cannot directly despawn it.
To address this, we employ Remote Procedure Calls (RPCs): other users (RpcSources.All) send despawn requests to the first client (RpcTargets.StateAuthority), which then executes the despawn action.
A similar process applies when multiple users fire bullets that affect the health of adversaries.
In contrast, dynamic events such as explosions occur probabilistically in fire zones.
Here, the logic is reversed: the first client (RpcSources.StateAuthority) generates the explosion event and broadcasts it to all participants (RpcTargets.All) so that every user experiences the same hostile environment.

The weapon system requires more work to maintain consistency.
For example, when users pass a gun, we must transfer StateAuthority to the receiving player to ensure that the appropriate player can operate the weapon.
The system also carefully synchronizes bullet management.
A player can reload only the gun for which they hold StateAuthority, but each client also tracks the remaining bullets in other players’ guns.
This lets all users maintain a consistent and accurate view of weapon states in the shared environment.

\textbf{Player-related states and mission metrics.}
Our platform maintains consistent player status information across all users to preserve shared situational awareness.
Each player’s health is visible as a health bar positioned above their head, which updates in real time.
We synchronize the six-degree-of-freedom (6DoF) head pose of each player to keep the health bar’s position and rotation accurate.
To ensure smooth coordination during the collaborative rescue task and elevator interaction, the platform also synchronizes player hand movements, replicating each user’s hand motions across clients to support seamless interaction.
Beyond individual states, the system synchronizes and displays team-level mission metrics on each player's HUD, such as the number of adversaries defeated and hostages rescued, which facilitates communication and collective progress tracking.
These synchronizations ensure that all users maintain awareness of both individual statuses and team achievements, enhancing the design and execution of effective collaborative strategies.

\section{Evaluation of Initial Synchronization}
\input{figure/11-sync}
Initial synchronization time refers to the initial process by which all users join the network room and achieve colocation.
We define it as a three-stage process:
(1) the first client connects to the Photon Fusion server, creates a network room, and initializes a synchronization anchor;
(2) the second client joins the room and the first client shares the anchor; and
(3) the second client loads the anchor locally.
Successful synchronization is confirmed when the anchor appears at the same location for both users.

As shown in~\Cref{fig:sync_latency}, the anchor sharing stage is the most time-consuming, with an average of 8.6s, since it involves colocation between two users.
The first client’s creation stage is the shortest, averaging 2.7s, while the second client’s loading stage averages 4.6s.
Importantly, this synchronization process occurs only once during session setup.
After all users are synchronized, subsequent synchronization times drop to minimal levels around 0.089s.

\section{Dataset Format}
\input{table/1-format}

The \sysname{} testbed exports all recorded data in a session-based directory structure to support reproducibility and post-hoc analysis. 
Each experiment session is stored as an individual folder containing synchronized logs of system events, user behaviors, subjective responses, and multimodal sensing data. 
Files are organized by modality so that different analysis pipelines can directly access task performance, cognitive attack events, and behavioral measurements.
The dataset includes system-level logs such as attack, interaction events, and task progression, as well as user-level records including questionnaire responses, head movement, and eye gaze data. 
Additional first-person video and audio recordings are stored alongside structured logs for contextual analysis. 
A summary of the main file categories is shown in~\Cref{tab:data_format}.

%% file: figure/04-HP.tex
\begin{figure}[t]
    \centering
    \includegraphics[width=0.47\textwidth]{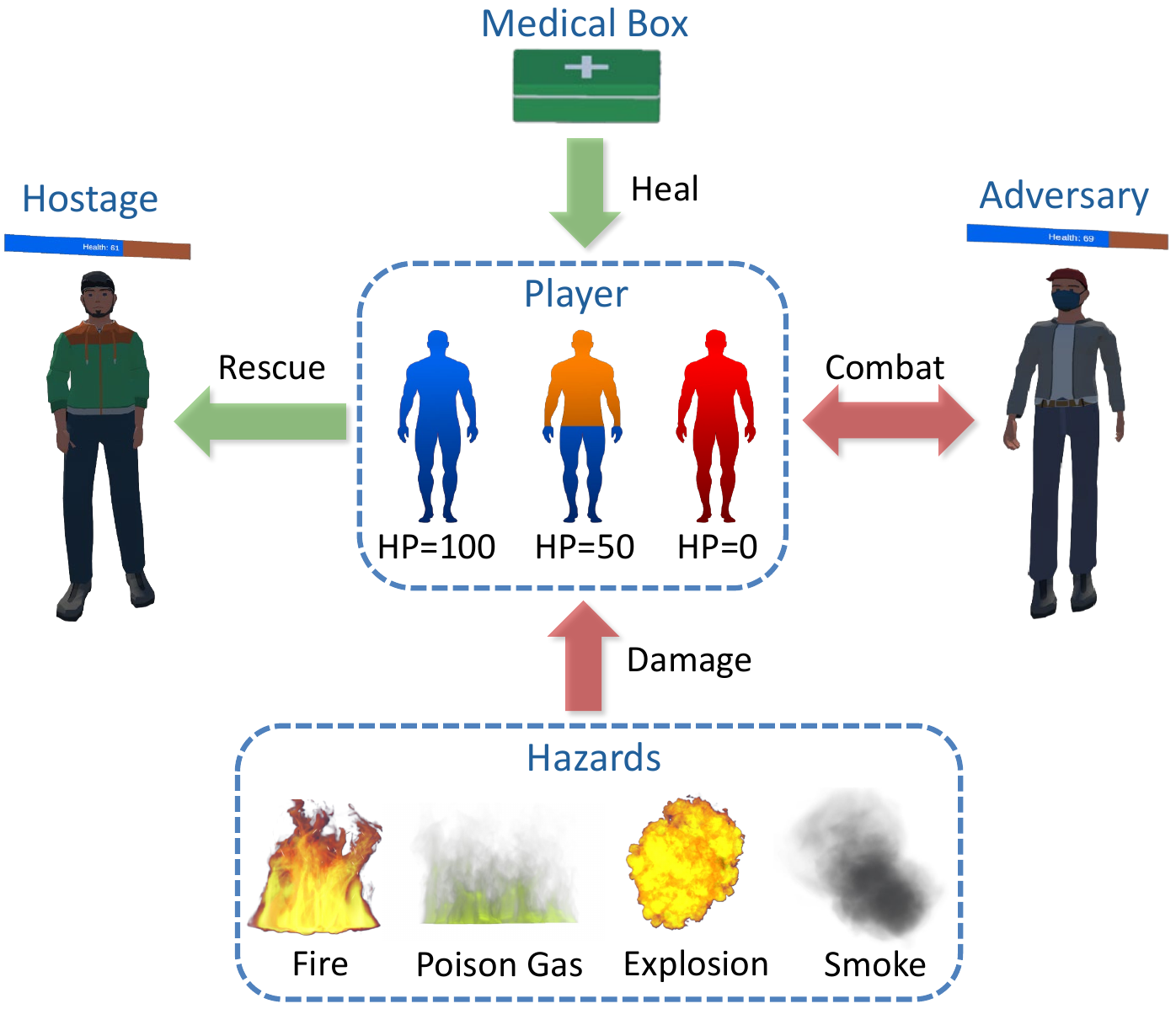}
     \caption{Health point (HP) framework. Players lose health points (HP) through combat with adversaries and exposure to hazards. Health can be restored using medical boxes, and players must maintain sufficient HP to successfully rescue hostages.}
    \label{fig:health}
\end{figure}

%% file: figure/05-NPC.tex
\begin{figure*}[tb]
    \includegraphics[width=0.32\textwidth]{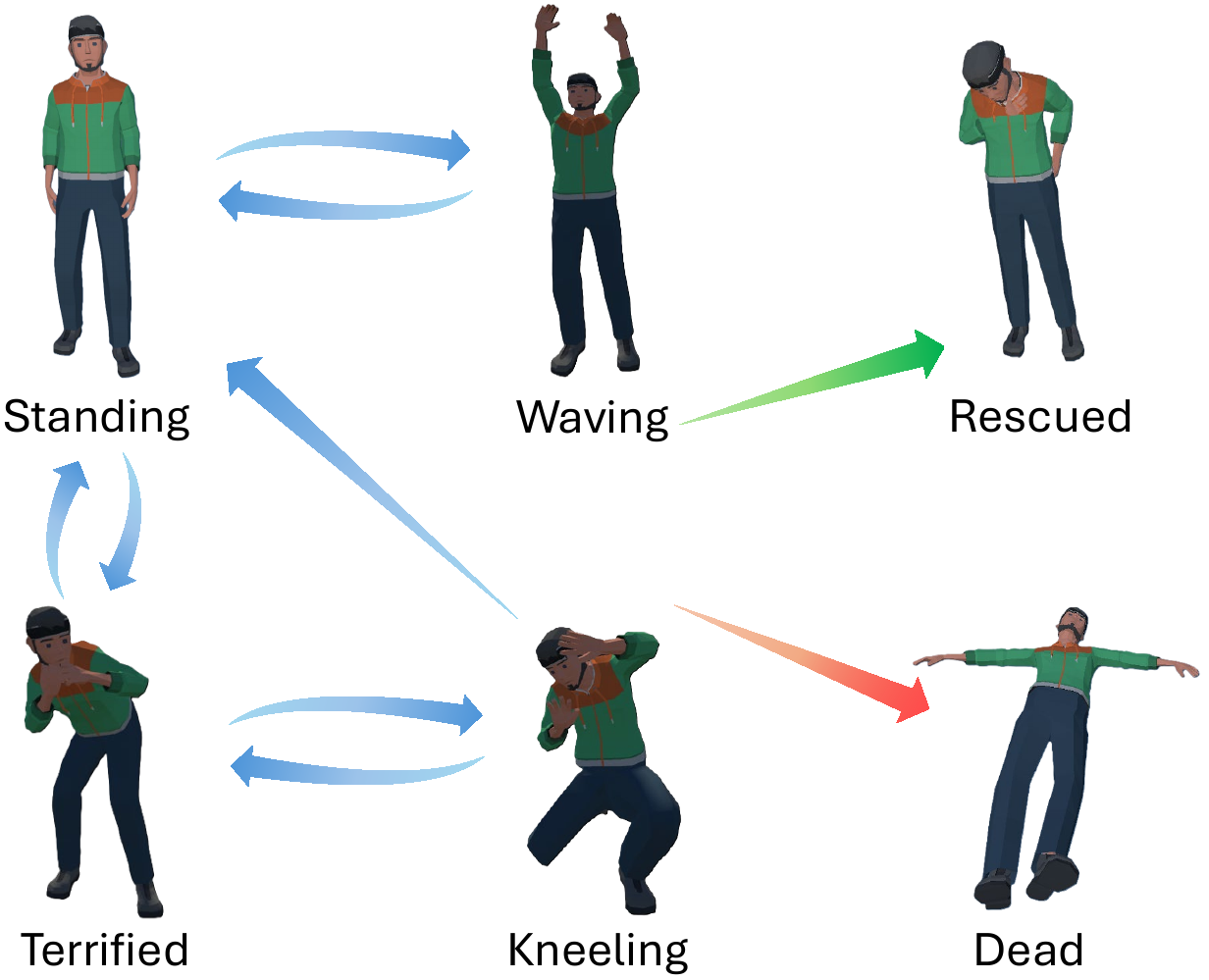}\label{fig:hostage_animation}
    \hfill
     \includegraphics[width=0.32\textwidth]{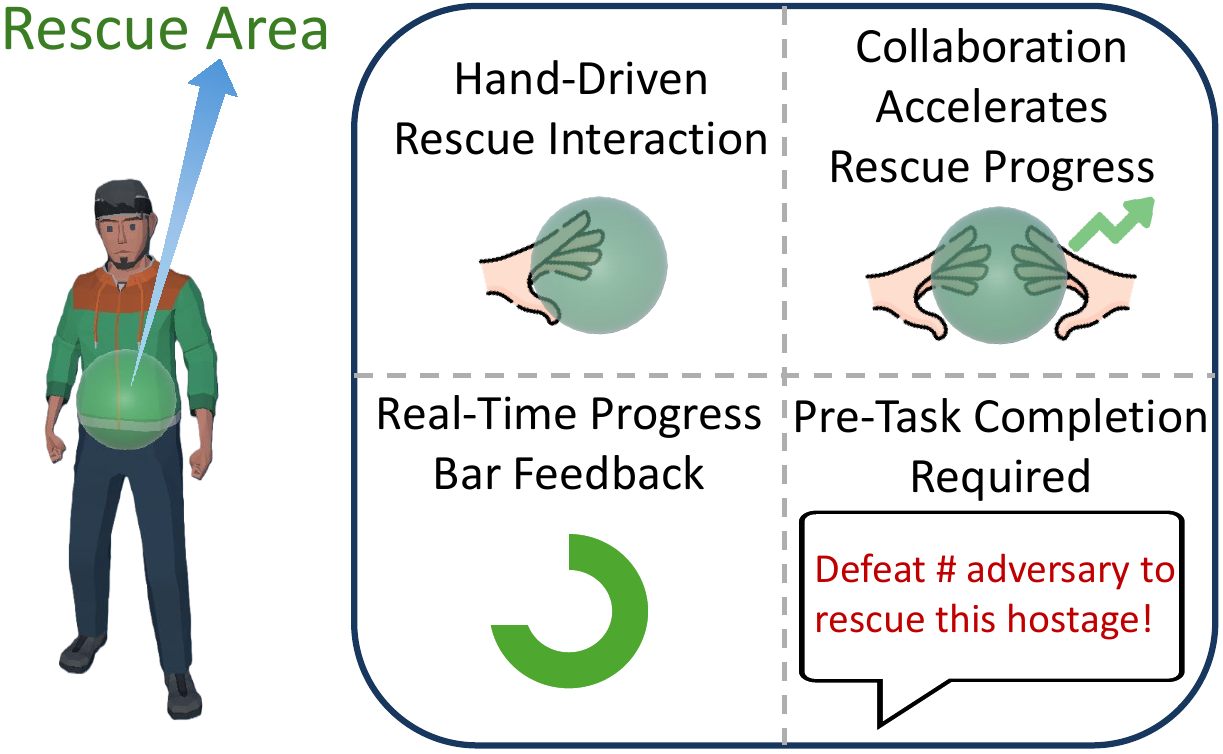}\label{fig:rescue_mechanism}
    \hfill
    \includegraphics[width=0.32\textwidth]{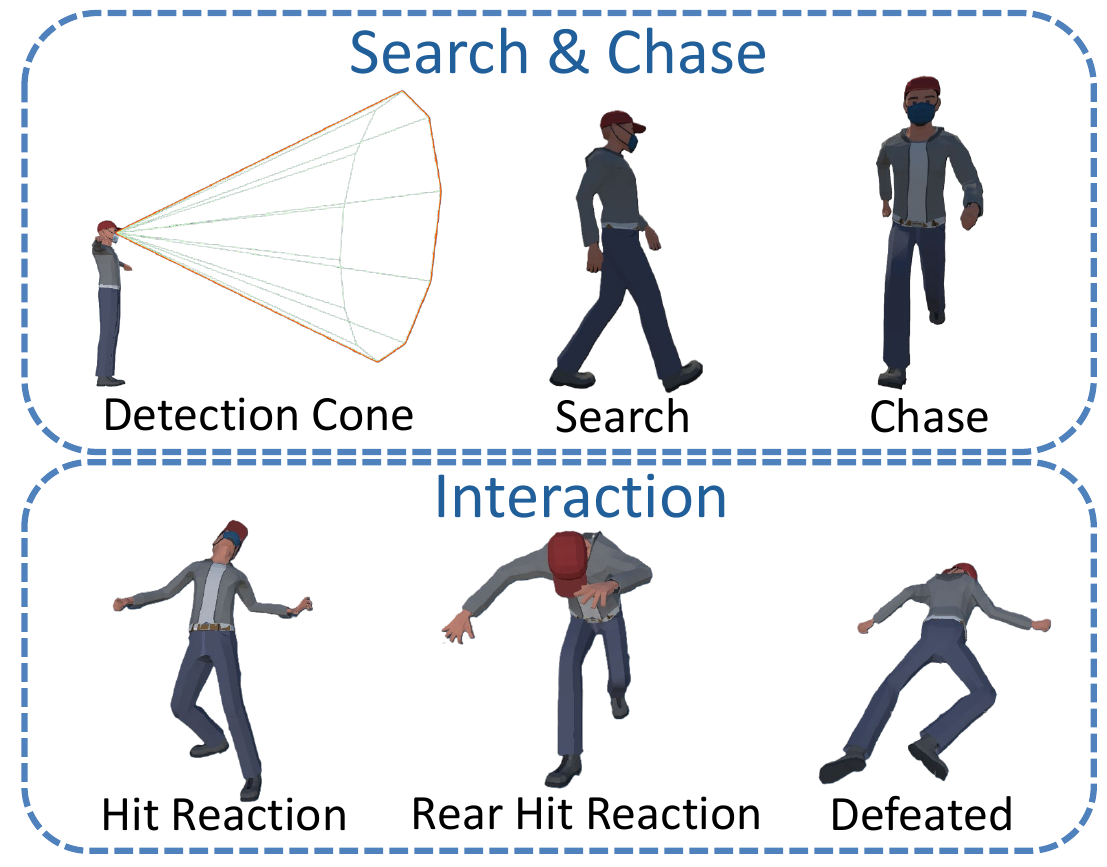}\label{fig:adversary}
    \caption{Hostage and adversary behaviors.
    (\textbf{Left}): Animation transitions among different hostage states.
    (\textbf{Center}): Rescue mechanism involving hand-driven interaction, progress feedback, collaboration, and pre-task completion check.
    (\textbf{Right}): Adversary AI's search, chase, and interaction animations.}
    \label{fig:npcs}
\end{figure*}

%% file: figure/06-weapon+.tex
\begin{figure}[tb]
    \begin{minipage}[b]{0.55\columnwidth}
    \includegraphics[width=\columnwidth]{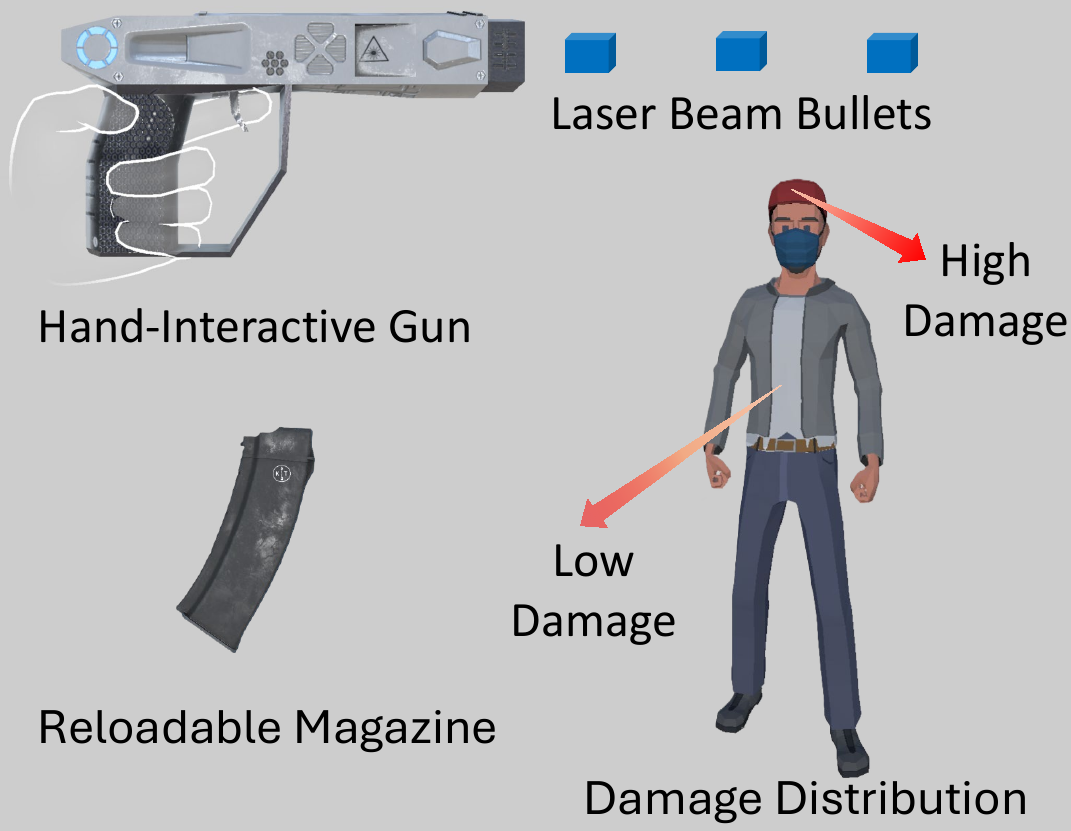}
    \end{minipage}
    \hfil
    \begin{minipage}[b]{0.39\columnwidth}
    \includegraphics[width=\columnwidth]{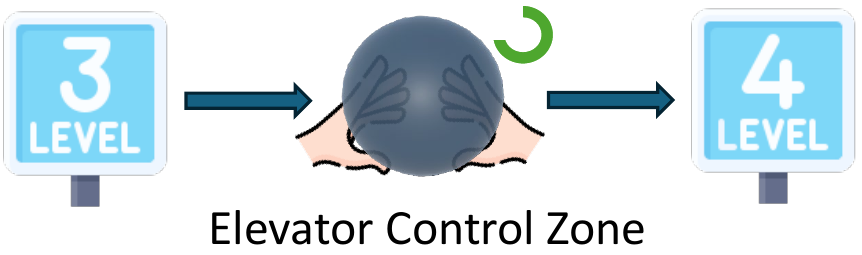}
    \\\vfill
	\includegraphics[width=\columnwidth]{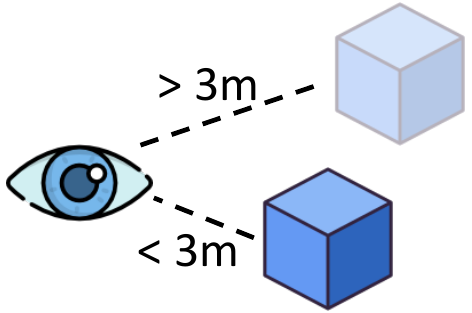}
    \end{minipage}
    
    \caption{Illustrations of game mechanics.
    (\textbf{Left}): Weapon system. The hand-interactive gun fires laser beam bullets using a reloadable magazine. Damage is distributed based on hit location.
    (\textbf{Top right}): Players' interaction with the elevator control area.
    (\textbf{Bottom right}): Distance-based visibility mechanics. This component reduces player visibility within a specified radius, for example, 3m.}
    \label{fig:weapon}
    
\end{figure}

%% file: figure/11-sync.tex
\begin{figure}[tbh]
    \centering
    \includegraphics[width=0.8\columnwidth]{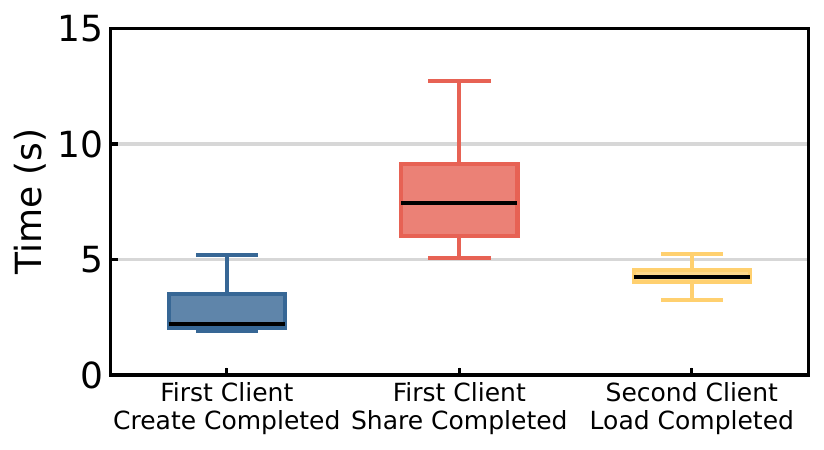}
    \caption{Client synchronization time over 50 samples.
    Left to right: processing time for the first client to connect to the server, time for the first client to upload the synchronization anchor to the server, and time for the second client to receive the anchor from the server.}
    \label{fig:sync_latency}
\end{figure}

%% file: table/1-format.tex
\begin{table*}[tbh]
\centering
\caption{Dataset format.}
\label{tab:data_format}
\begin{tabular}{@{}m{0.2\columnwidth} | m{1.5\columnwidth{}}@{}}
\hline
\textbf{Cognitive \newline Attack} & 
Date-Time: \texttt{MM/DD/YY HH:MM:SS.sss}, e.g., ``12/01/2025 10:46:17.509''
\newline
Timestamp: \texttt{MM:SS.sss}, e.g., ``12:34.567'' \newline
Current-Floor: \texttt{int}
\newline
AttackType: \texttt{string} (\textit{attack\_type}), e.g., ``Latency'' \newline
Intensity: \texttt{int} \\
\hline

\textbf{User \newline Subjective \newline Measures} & 
Date-Time: \texttt{MM/DD/YY HH:MM:SS.sss}
\newline
Timestamp: \texttt{MM:SS.sss} \newline
Current-Floor: \texttt{int}
\newline
EngagementQ1Score, EngagementQ2Score, EngagementQ3Score, EngagementQ4Score: \texttt{int} \newline
TrustQ1Score, TrustQ2Score, TrustQ3Score, TrustQ4Score: \texttt{int} \\
\hline

\textbf{Behavioral \newline Data}
 & 
Date-Time: \texttt{MM/DD/YY HH:MM:SS.sss}
\newline
Timestamp: \texttt{MM:SS.sss} \newline
Current-Floor: \texttt{int}
\newline
\textit{Eye Tracking:} \newline
-- LeftEyePositionX/Y/Z: \texttt{float} \newline
-- LeftEyeRotationX/Y/Z/W: \texttt{float} \newline
-- LeftGazeDirectionX/Y/Z: \texttt{float} \newline
-- LeftGazePoint2DX/2DY: \texttt{float} \newline
-- RightEyePositionX/Y/Z: \texttt{float} \newline
-- RightEyeRotationX/Y/Z/W: \texttt{float} \newline
-- RightGazeDirectionX/Y/Z: \texttt{float} \newline
-- RightGazePoint2DX/2DY: \texttt{float} \newline
GazeTarget: \texttt{string} (\textit{obj\_name}), e.g., ``Hostage'' \newline
ObjectInfo: \{ObjName: \texttt{string}, PixelPosition: \{x: \texttt{float}, y: \texttt{float}\}\}, e.g., \{`ObjName': ``Fire'', `PixelPosition': (1924.83, 278.37)\} \newline
ObjCount: \texttt{int} \newline
\textit{Head Tracking:} \newline
-- HeadPositionX/Y/Z: \texttt{float} \newline
-- HeadRotationX/Y/Z/W: \texttt{float} \\
\hline

\textbf{User \newline Activity \newline Event} &
Date-Time: \texttt{MM/DD/YY HH:MM:SS.sss}
\newline
Timestamp: \texttt{MM:SS.sss} \newline
Current-Floor: \texttt{int}
\newline
EventType: \texttt{string} (\textit{event\_type}), e.g., ``HostageRescued'' \newline
EventDescription: \texttt{string} (\textit{event\_description}), e.g., ``Hostage ID: 5. HostageHealthPoint: 19. TotalHostageSaved: 4'' \\
\hline
\end{tabular}
\end{table*}